\documentclass[fontsize=10pt,paper=a4,bibliography=totoc,headsepline=true
,abstract=true]{scrartcl}

\usepackage[automark]{scrlayer-scrpage}
  \clearpairofpagestyles
  \ihead[]{\headmark}
\usepackage[inner=2.5cm, outer=2.5cm, top=3cm, bottom=3cm]{geometry} 

\usepackage[T1]{fontenc}
\usepackage[utf8]{inputenc}
\usepackage{lmodern}                    
\usepackage[british]{babel} 
\usepackage[colorlinks=true]{hyperref}

\usepackage[auth-sc-lg, noblocks]{authblk}

\usepackage{csquotes}
\usepackage[style=numeric-comp,doi=true,url=true,maxbibnames=5,backend=biber
,sorting=none]{biblatex}
\bibliography{../literature}

\usepackage{amssymb, amsmath, amsthm}
\usepackage{thmtools}
\usepackage{mathtools}
\usepackage{commath}      
\usepackage{nicefrac}
\usepackage{array}  
\usepackage{multirow}

\usepackage{bm}
\usepackage{dsfont}
\usepackage{calrsfs}
\usepackage{stmaryrd}
\DeclareMathAlphabet{\pazocal}{OMS}{zplm}{m}{n}
\DeclareMathAlphabet{\pazocalbf}{OMS}{cmsy}{b}{n}
\usepackage{etoolbox}
\preto\subequations{\ifhmode\unskip\fi}

\usepackage{graphicx} 
\usepackage[export]{adjustbox}
\usepackage[dvipsnames]{xcolor}

\usepackage{enumitem}
  \setlist[itemize]{nosep, topsep=0pt, wide = 1em, leftmargin=*}  
\usepackage{booktabs}

\usepackage{X/siunitx}
\setkomafont{author}{\scshape}
\setkomafont{section}{\large}
\setkomafont{subsection}{\normalsize}
\setkomafont{subsubsection}{\normalsize}
\setkomafont{paragraph}{\normalsize}
\setkomafont{caption}{\sffamily\small}
\setkomafont{captionlabel}{\bfseries\usekomafont{caption}}

\definecolor{darkred}{RGB}{139,0,0}
\definecolor{mediumblue}{RGB}{0,0,205}
\definecolor{forestgreen}{RGB}{34,139,34}
\hypersetup{linktocpage=true,linkcolor=darkred,urlcolor=forestgreen,citecolor
=mediumblue}

\declaretheoremstyle[
  spaceabove=\topsep, spacebelow=\topsep,
  headfont=\normalfont\bfseries,
  notefont=\normalfont\bfseries, notebraces={(}{)},
  bodyfont=\normalfont,
  postheadspace=1em,
  qed=$\blacktriangle$,]{myremstyle}

\declaretheorem[style=myremstyle]{example}
\declaretheorem[style=myremstyle, numbered=no]{remark}
\declaretheorem[style=myremstyle, name=Remark, sibling=example]{remnum}

\numberwithin{equation}{section}
\AtBeginEnvironment{tabular}{\small}

\newcommand{\R}{\mathds{R}}							
\newcommand{\PP}{\mathbb{P}}
					
\renewcommand{\L}[2]{\pazocalbf{L}^{#1}(#2)}		
\newcommand{\Ltwozero}[1]{\pazocal{L}^2_0(#1)}		
\renewcommand{\H}[3]{\pazocalbf{H}^{#1}_{#2}(#3)}	
\newcommand{\nrm}[2]{\Vert#1\Vert_{#2}}				

\renewcommand{\O}{\Omega}							

\newcommand{\OO}{\pazocal{O}}								
\newcommand{\FL}{\pazocal{F}}						
\newcommand{\SO}{\pazocal{S}}						
\newcommand{\IN}{\pazocal{I}}						
\newcommand{\BO}{\pazocal{B}}                       

\newcommand{\V}{\bm{V}}								
\newcommand{\THk}[1]{\texttt{TH}_{#1}}

\newcommand{\ub}{\bm{u}}

\newcommand{\n}{\bm{n}}								
\newcommand{\x}{\bm{x}}								
\newcommand{\f}{\bm{f}}								
\newcommand{\restr}[2]{{\left.\kern-\nulldelimiterspace#1\right|_{#2}}}	

\DeclareMathOperator{\Id}{Id}						
\DeclareMathOperator{\diver}{div}					
\renewcommand{\div}{\diver}
\DeclareMathOperator{\vol}{vol}

\newcommand{\stress}{\bm{\sigma}(\ub,p)}

\newcommand{\nf}[2]{\nicefrac{#1}{#2}}
\newcommand{\dt}{\Delta t}							
\newcommand{\st}{\;\vert\;}

\begin{document}

\title{\Large Using a deep neural network to predict the motion of
under-resolved triangular rigid bodies in an incompressible flow}

\author[a,$\ast$]{Henry von Wahl}
\author[a]{Thomas Richter}

\affil[a]{Institute for Analysis and Numerics, Otto-von-Guericke Universität,
Universitätsplatz~2, 39106~Magdeburg, Germany}
\affil[$\ast$]{Corresponding author: \texttt{henry.vonwahl@ovgu.de}}

\date{\small\today}
\maketitle

\begin{abstract}
We consider non-spherical rigid body particles in an incompressible fluid in
the regime where the particles are too large to assume that they are simply
transported with the fluid without back-coupling and where the particles are
also too small to make fully resolved direct numerical simulations feasible. 
Unfitted finite element methods with ghost-penalty stabilisation are well
suited to fluid-structure-interaction problems as posed by this setting, due to
the flexible and accurate geometry handling and for allowing topology changes
in the geometry. 
In the computationally under resolved setting posed here, accurate computations
of the forces by their boundary integral formulation are not viable.
Furthermore, analytical laws are not available due to the shape of the
particles. 
However, accurate values of the forces are essential for realistic motion of
the particles.
To obtain these forces accurately, we train an artificial deep neural network
using data from prototypical resolved simulations.
This network is then able to predict the force values based on information
which can be obtained accurately in an under-resolved setting. As a result, we
obtain forces on very coarse and under-resolved meshes which are on
average an order of magnitude more accurate compared to the direct
boundary-integral computation from the Navier-Stokes solution, leading
to solid motion
comparable to that obtained on highly resolved meshes that would
substantially increase the simulation costs.
\end{abstract}

\section{Introduction}
\label{sec.intro}

The aim of this work is to simulate multiple small triangular rigid bodies
inside an incompressible fluid using a moving domain approach \cite{LO19,
BFM19, vWRL20}, within an
unfitted finite element method known as CutFEM \cite{BCH+14}.
Solids in flows are the subject of various applications in engineering
and medicine. The homogenised behaviour of these particles is well
studied, but numerical simulations on the fine scale, considering
resolved particles are rare, especially when non-spherical bodies are
considered.

If the solid particles within the fluid are very small, they can be
considered as point-like within the flow field and the interaction can be
represented in a discrete element framework in terms of analytical laws for
the fluid mechanical coefficients, as well as by modelling the effect
of the bodies on the surrounding flow. Approximative laws exist for
some simple shaped non-spherical
particles~\cite{F94,HS08,MR10,ZMZvW12,ZYL16} and simulations with a very
large number of particles can be realised in mixed continuum
mechanics / discrete element frameworks.

However, if the particles are assumed to be large, only a resolved
simulation remains. When considering spherical particles, Arbitrary Lagrangian
Eulerian (ALE) approaches can be used. These allow for very good accuracy on
relatively coarse grids~\cite{WT07,vWRFH21}. 
The ALE method is based on fixed
reference domains that allow to resolve the boundaries of the particles
exactly. The motion of the particles is implicitly tracked by mapping the flow
problem onto this reference domain. ALE approaches are highly efficient and
accurate but whenever the motion (or rotation) of the particles gets too large,
they suffer from a distortion of the underlying finite element mesh and call
for remeshing.
This approach is highly efficient and very well
established in the area of elastic fluid structure
interactions~\cite{HLZ81,D82,RW10}. However, if a very large number of
particles are  considered, or if they are no longer radially
symmetric, alternative discretisation methods must be considered as
tracking the single particles with a finite element mesh may require a
prohibitive computational effort. A prominent example is the immersed
boundary method, going back to Peskin~\cite{Pes02}, which is able to
efficiently simulate dense particle
suspensions~\cite{AEvW18}. Although the particles must not be exactly
resolved, these methods still require an increased mesh resolution in
the vicinity of the particles. 

Finally, we note that on a theoretical level, the motion of rigid bodies of
arbitrary shapes has been studied by Brenner and summarised by Happel and
Brenner in \cite{HB81}. However, these considerations only take into account
the case of creeping flows governed by the linear Stokes equations,
rather than the full non-linear Navier-Stokes equations considered here.

In this work we define a hybrid finite element / (deep)
neural network framework that allows us to describe the interaction
with rigid bodies of triangular shape on
under-resolved meshes accurately. Triangular bodies are a simple
example of non-spherical shapes where no rotational symmetry can be assumed
that would allow us to use efficient and accurate ALE approaches. Furthermore,
since triangles involve re-entrant corners, the approximation property of the
finite element solution suffers, which poses another challenge to the coupled
simulation. Both problems also arise for general polygonal objects, but by
choosing triangles we can restrict ourselves to geometries that are easy to
parametrise in this demonstration. In order to write the transfer
of forces between flow and particles with sufficient accuracy, we will
use a neural network that represents the forces acting on a particle
as a function  of the particular flow situation instead of directly
evaluating the forces from the Navier-Stokes solution. This network is
trained in a presented offline phase based on prototypical resolved
simulations. A similar approach with non-spherical but very small 
particles in the linear Stokes regime has recently been described by
Minakowska et al.~\cite{MRS20}. In principle this procedure can be
transferred to all interface capturing schemes that allow for a robust
under-resolved representation of structures in a background
mesh. Besides different CutFEM or extended finite element
approaches~\cite{HH02,FD07,BE09,BCH+14} locally adjusted or parametric
finite element methods~\cite{LY05,F13,FR14} are also an option.  

Finally, we will illustrate that our neural network approach can
efficiently represent the
interaction of a fluid with non-spherical particles with high accuracy on
coarse grids. To date, there is no efficient alternative approach. ALE methods
are too costly due the the constant need for remeshing and resolved CutFEM
approaches are not competitive for their limited approximation properties that
would require highly refined meshes at the particles.

\begin{example}[Motivation of approach]
\label{ex.motivaton}
The approach is based in the hypothesis that while it may not be
possible to evaluate transmission forces with sufficient accuracy on
coarse and efficient discretisations, averaged flow features can
indeed be obtained obtained accurately on such meshes and that these
features are sufficient to predict the transmission forces with help
of a neural network. 
To illustrate that we can accurately compute volumetric flow features 
accurately, while the boundary forces are not computed accurately
in an under-resolved CutFEM simulations, we consider the well-established  
benchmark \emph{flow around a cylinder} \cite{ST96}, where the
laminar flow around a cylinder is considered and where the forces on
the cylinder are taken as goal values for a quantitative evaluation
of different discretisations, see \autoref{fig.motivation_example}
for a sketch of the solution. We note that the configuration is
slightly non-symmetric with the cylinder not being vertically
centred such that a non-vanishing lift coefficient will result.
Here we take the stationary "2D-1" test case and compute this once using a
fitted approach together with the Babu\v{s}ka-Miller trick \cite{BM84} to
evaluate the drag and lift functional, and once using a CutFEM approach
\cite{BH14, MLLR14} together with an isoparametric mapping approach
\cite{Leh16, Leh17} to obtain the necessary geometry approximation of the
discrete level set domain but with the direct evaluation of the boundary
integral to realise the force values. 
For both the fitted and unfitted simulations we use inf-sup stable Taylor-Hood
elements $\PP^k/\PP^{k-1}$, which we shall abbreviate as $\THk{k}$. For the
present computations we take the order $k=3$. To make the comparison as fair as
possible, we consider unstructured meshes with the same mesh size in each
computation.

These and all following numerical finite element computations are implemented 
using the finite element library
\texttt{NGSolve}/\texttt{netgen}~\cite{Sch14,Sch97}. For CutFEM computations,
we additionally use the add-on 
\texttt{ngsxfem}~\cite{ngsxfem} for unfitted finite element functionality
within the \texttt{NGSolve} framework.

In \autoref{fig.motivation_example} we can see the velocity solution to these
computations together with the computational meshes. Note that it is
nearly impossible to distinguish the two solutions visually. In
\autoref{tab.motivation-benchmark-quants} we see the benchmark
quantities resulting from the two computations. Here we immediately see that
while the values from the fitted computations are reasonably for
such coarse meshes, the values resulting from the CutFEM computations even
result in the wrong sign for the lift coefficient.

Since the solution in the bulk of the domains look so indistinguishable, it is
natural to ask, whether there are other quantities based on the solution near
the obstacle, which we can compute accurately in both settings. To this end, we
compute the average velocity in a strip of width $0.05$ around the obstacle.
The results from this are presented in \autoref{tab.motivation.avg}.
Here we see that while the fitted FEM solution resulted in values closer to the
reference values compared to CutFEM, the difference between the accuracy of the
fitted and unfitted computations are significantly smaller. Furthermore we see
that we keep multiple significant figures of accuracy in the functional value,
even at values of order $10^{-5}$.

We conclude from this, that while it is difficult to obtain accurate forces
from the boundary integral formulation in an under-resolved CutFEM computation,
other features of the solution can be obtained much more accurately even on
such coarse meshes. Therefore, if we are able to construct a mapping from flow
features near the obstacle to the forces acting on the obstacle, we should be
able to get more accurate force values in the under resolved CutFEM setting.

\begin{figure}
  \centering
  \includegraphics[width=.8\textwidth]{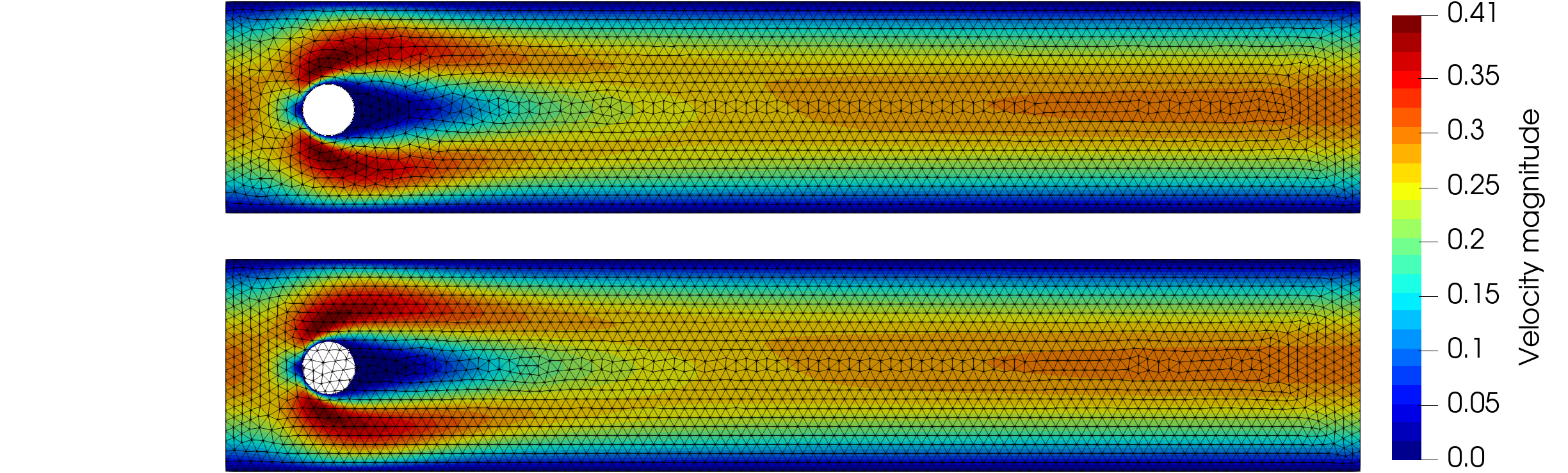}
  \caption{Velocity solution to the ST-2D1 problem with
    $h_{max}=0.02$. Top: Fitted FEM, bottom: CutFEM.}
  \label{fig.motivation_example}
\end{figure}

\begin{table}
  \centering
    \begin{tabular}{r llrlll}
      \toprule
      Method & $c_\text{drag}$& (err) & \multicolumn{1}{l}{$c_\text{lift}$}
      & (err) & $\Delta p$ & (err)\\
      \midrule
      Fitted $\THk{3}$ & 5.579521 & 0.00025\% &$ 0.0105942$&0.232\% &
      0.117162& 0.3046\%\\
      CutFEM $\THk{3}$ & 5.571512 & 0.14379\% &$-0.0063662$&160.0\% & 0.119766& 1.9111\%\\
      \cmidrule(lr){2-3}      \cmidrule(lr){4-5}      \cmidrule(lr){6-7}
      Ref. & 5.579535&-- & 0.0106189&-- & 0.117520&--\\
      \bottomrule
    \end{tabular}
    \caption[Benchmark quantities for the ST-2D1 problem computed using
      fitted FEM and CutFEM on meshes with $h_{max}=0.02$.]{Benchmark
      quantities for the ST-2D1 problem computed using fitted FEM and CutFEM on
      meshes with $h_{max}=0.02$. Reference values obtained from
\href{http://www.featflow.de/en/benchmarks/cfdbenchmarking/flow/dfg_benchmark1%
_re20.html}{\texttt{www.featflow.de}} (accessed on 15th Feb.\ 2021). }
    \label{tab.motivation-benchmark-quants}
\end{table}

\begin{table}
  \centering
  \sisetup{
    round-mode = figures,
    round-precision = 6,
    table-format=1.5e1,
    scientific-notation=true,
    exponent-product = \cdot,
  }
  \begin{tabular}{r llll}
    \toprule
    Method & Avg. $\ub_1$ & (err) & Avg. $\ub_2$ & (err)\\
    \midrule
    Fitted $\THk{3}$ & \num{0.004293270580011} & 0.0013\% &
    \num{1.49200245778517e-05} & 0.0306\%\\
    CutFEM $\THk{3}$ & \num{0.004287956263878} &1.2506\%  &
    \num{1.4931650464672e-05} & 0.1086\%\\
    \cmidrule(lr){2-3}    \cmidrule(lr){4-5}
    Ref. &  \num{0.0042933255956392} & -- & \num{1.49154542267073e-05}
    & --\\
    \bottomrule
  \end{tabular}
  \caption{Average velocity in a strip of width 0.05 around the obstacle.
    Reference values computed using a fitted approach with $\THk{6}$ elements
    on a mesh with $h_{max}=0.005$. We also indicate the relative
    error of both approaches compared to the reference solution. }
  \label{tab.motivation.avg}
\end{table}
\end{example}

The remainder of this text is structured as follows. In the following
\autoref{sec.equations} we describe the basic equations governing the system of
a rigid body in an incompressible fluid. In
\autoref{sec.equations:subsec.param} we then discuss the relevant physical
parameters we will consider. \hyperref[sec.network]{Section~\ref*{sec.network}}
covers the details of the neural network, that is the design, the generation of
the training data and the results of the training. Furthermore, we
validate the accuracy of the network prediction in comparison to the
direct evaluation from
the Navier-Stokes solution in the setting of the generated training data.
In \autoref{sec.num} we will then apply the resulting network in different
scenarios, in order to compare the results against a resolved ALE solution and to
show the capabilities of the method in settings where a standard ALE
discretisation is not realisable. In
\autoref{sec.conclusion} we give some concluding remarks.

\section{Governing equations}
\label{sec.equations}

Let us consider an open bounded domain $\O = \FL\:\dot\cup\:\IN\:\dot\cup\:\SO
\subset\R^{d}$, $d\in\{2,3\}$, divided into a $d$-dimensional fluid region
$\FL$, a $d$-dimensional solid region $\SO$ and a $d-1$-dimensional interface
region $\IN$ coupling the two. The fluid in $\FL$ is governed by the 
incompressible Navier-Stokes equations: Find a velocity and pressure $(\ub, p)$
such that
\begin{equation}\label{eqn.navier-stokes}
  \begin{aligned}
    \rho_\FL\left(\partial_t\ub + (\ub\cdot\nabla)\ub\right) - \div\stress
      &=\rho_\FL\f \\
    \div(\ub) &= 0
  \end{aligned} 
\end{equation}
holds in $\FL$ with a given fluid density $\rho_\FL$, an external body force
$\f$ and the Cauchy stress tensor
\begin{equation*}\stress = \mu_\FL(\nabla\ub + \nabla\ub^T) - \Id p
\end{equation*}
where $\Id$ is the identity tensor, $\mu_\FL=\rho_\FL\nu_f$ is the fluid's 
dynamic viscosity and $\nu_\FL$ the kinematic viscosity.
The boundary conditions which complete the system
\eqref{eqn.navier-stokes} will be given later. We shall consider the
homogeneous form of \eqref{eqn.navier-stokes}, i.e., we take $\f\equiv 0$.

We further divide the solid region $\SO = \dot\bigcup_{i=1}^N \BO_i$ into a
finite set of distinct rigid bodies (particles) $\BO_i$. For simplicity, we
assume that they each have the same homogeneous density $\rho_\SO$. Let
\begin{equation*}
  U_i^\text{total} = U_i +\bm{\omega}_i\times\bm{r}
\end{equation*}
be the motion of $\BO_i$, where $U_i$ is the
particle's \emph{velocity}, $\bm{\omega}_i$ it's \emph{angular velocity} and 
$\bm{r}=\x- c_\BO$ the position vector of a point in $\overline{\BO}_i$
relative to the body's centroid $c_\BO$. These two velocities are then governed
by the Newton-Euler equations
\begin{subequations}\label{eqn.solid-ode}
  \begin{align}
    m_\BO\partial_t U_i 
      &= \bm{F} + \bm{F}_\text{buoyant} + \bm{F}_\text{gravity}\\
    \bm{I}_\BO\partial_t\bm{\omega} +\bm{\omega}\times\bm{I}_\BO\bm{\omega} 
      &= \bm{T} + \bm{T}_\text{buoyant}\label{eqn.solid-ode:rot}
  \end{align}
\end{subequations}
with the particles mass $m_\BO = \rho_\SO\vol(\BO)$, the force and torque
exerted by the fluid on the particle
\begin{equation*}\bm{F}=\int_{\partial\BO_i}\stress\n\dif S
  \qquad\text{and}\qquad
  \bm{T} = \int_{\partial\BO_i} \bm{r}\times\stress \n \dif S
\end{equation*}
and the particles moment of inertia tensor defined by
\begin{equation*}\bm{I}_\BO = \rho_\SO \int_{\BO_i}(\nrm{\bm{r}}{}^2\Id_3 -
    \bm{r}\otimes\bm{r})\dif\x.
\end{equation*}
The buoyancy force and torque are
\begin{equation*}
  \bm{F}_\text{buoyant} = -m_\FL \bm{g}
  \quad\text{and}\quad
  \bm{T}_\text{buoyant} = -m_\FL \bm{r}_{bo}\times \bm{g}
\end{equation*}
where $m_\FL = \rho_\FL\vol(\BO_i)$ is the mass of the displaced fluid and
$\bm{r}_{bo}$ is the vector from the centre of mass to the centre of buoyancy.
The centre of buoyancy is defined as the centroid of the displaced fluid
volume. Since the particles are completely submersed and both the fluid and 
solid have a constant density, the centre of mass and centre of buoyancy 
coincide, such that $\bm{r}_{bo} = 0$ and therefore  $\bm{T}_\text{buoyant}=0$.
Note that equivalently, the buoyancy effects can be included in the system by
including the forces due to gravity on the right hand side of
\eqref{eqn.navier-stokes}. However, since we do not consider pressure-robust
discretisations \cite{JLM+17} here, including buoyancy in the solid and
ignoring gravity on the fluid is more accurate on the discrete level. 
Finally, the pull due to gravity is
\begin{equation*}
  \bm{F}_\text{gravity} = m_\SO\bm{g}.
\end{equation*}

To complete the system \eqref{eqn.navier-stokes} we need to impose boundary
conditions. We shall consider $\O$ as a closed aquarium, and therefore take
the no-slip boundary conditions
\begin{equation}\label{eqn.navier-stokesBCs}
   \ub =\bm{0} \text{ on }\partial\O
   \qquad\text{and}\qquad
   \ub = U_i^\text{total} \text{ on }\partial\BO_i
\end{equation}
which couples the fluid and solid equations. The complete system is then
defined as the solution to \eqref{eqn.navier-stokes}, \eqref{eqn.solid-ode} and
\eqref{eqn.navier-stokesBCs}. The pressure is uniquely defined by taking
$p\in\Ltwozero{\FL}$.

\begin{remark}
We note that in two spatial dimensions, the quadratic term
$\bm{\omega}\times\bm{I}_\BO\bm{\omega}$ vanishes in \eqref{eqn.solid-ode:rot}
and the moment of inertia reduces to the scalar
$I_\BO = \rho_\SO\int_\BO \nrm{\bm{r}}{}^2\dif\x.$
\end{remark}

\subsection{Choice of parameters}
\label{sec.equations:subsec.param}

We choose our fluid and solid material parameters, in order to have a setting
which can be given some physical meaning while remaining at small to moderate
Reynolds numbers.

The fluid and solid parameters are chosen to approximate coarse sand in
a glycerol/water mixture. We take a mixture of 1 part water to 4 parts
glycerol at a temperature of $21\unit{\degreeCelsius}$. The resulting relevant 
material parameters are summarised in \autoref{tab.parameters}. The fluid 
parameters are obtained through an online calculator tool\footnote{
\url{http://www.met.reading.ac.uk/~sws04cdw/viscosity_calc.html}, accessed on
25\textsuperscript{th} Sep.\ 2020} and the density of the solid is taken as the
density of quartz\footnote{
\url{http://www.matweb.com/search/datasheet_print.aspx?matguid
=8715a9d3d1a149babe853b465c79f73e}, accessed on 25\textsuperscript{th} Sep.\
2020}.
The ISO standard 14688-1:2017\footnote{%
\url{https://www.sis.se/api/document/preview/80000191}, 
accessed on 25\textsuperscript{th} Sep.\ 2020}
defines coarse sand to have a particle size of 0.63 -- 2.0 \si{\milli\m}. We
therefore take the side length of our triangles to be
$d_\BO=\SI{2e-3}{\metre}$.

\begin{table}[!ht]
  \centering
  \begin{tabular}{r cccc}
    \toprule
    Parameter & $\mu_\FL$ & $\rho_\FL$ & $\nu_\FL$ & $\rho_\SO$\\
    \midrule
    Value & \SI[round-precision=4]{0.085679}{\newton\s\per\m\squared}
      & \SI[round-precision=4]{1216.7}{\kg\per\m\cubed}
      & \SI[round-precision=4]{0.000070419}{\m\squared\per\s}
     & \SI[round-precision=2]{2650}{\kg\per\m\cubed}\\
    \bottomrule
  \end{tabular}
  \caption{Fluid and solid material properties of a 1:4 water-glycerine 
    mixture and quartz}
  \label{tab.parameters}
\end{table}

With these parameters, we have established the terminal settling velocity of a
particle without rotational or horizontal motion to be $v=0.047\si{\m\per\s}$,
c.f. \autoref{sec.network:subsec.data:subsubsec.apriori}.
Taking the side-length of the triangular particle as the reference length,
this leads to a Reynolds number of $Re=\frac{vL}{\nu_\FL}\approx1.33$.
This is sufficiently large to justify considering the full Navier-Stokes
equations, rather than the creeping flow Stokes equations, and small enough to
ensure that the local flow around the triangle is not turbulent.

\section{Neural Network}
\label{sec.network}

As we have discussed above, our aim is to train a (deep) neural network which
predicts the forces acting on a small triangular rigid body, based on 
information retrieved from the fluid solution within the bulk of the fluid near
the obstacle. In principle this should be possible, for example Raissi 
et.\ al.~\cite{RYK20} where able to reproduce the Navier-Stokes solution from
images using a physics informed neural network~\cite{RPK19} or Minakowska et.\ 
al.~\cite{MRS20} used a very simple deep neural network to predict the forces 
acting on blood platelets of different shapes in a Stokes flow.

\subsection{Network design}
\label{sec.network:subsec.design}

As the work in~\cite{MRS20} is relatively similar with respect to the aim we
have here, we
shall take this as our reference point for the design of our network. We note
that while in~\cite{MRS20} it was the aim to use a neural network to predict
the forces based on the speed and shape of very small particle in a linear
Stokes flow and without back-coupling of the particles onto the fluid, we want
to predict the forces based on the fluid solution near our particle which
couples back to the fluid which is governed by the non-linear Navier-Stokes
equations.

\subsubsection{Network Input}
\label{sec.network:subsec.design:subsubsec.input}
The first question we have to consider is the choice of flow features to feed 
into the network. To capture the force acting on the triangular rigid body, it 
makes sense that the features have to be in some sense local to the rigid 
body. Point evaluations of the velocity and pressure near the rigid body 
would be one choice. Unfortunately, while we found that this does indeed 
capture the necessary information needed by a neural network, the unresolved 
nature of the CutFEM discretisation means that we do not have a chance of 
obtaining sufficiently accurate values to feed to the network at run-time.

As we have seen in \autoref{ex.motivaton}, an integral of the velocity
components, can be obtained accurately in a coarse CutFEM computation. 
For a rigid body $\BO$, we define
$\OO(\BO) \coloneqq \{ \x\in\FL \st \nrm{\x-c_\BO}{2}^2 < (d_\OO/2)^2 \}$ 
to be a circular fluid domain around the solid, centred at the solid's centre
of mass with radius $d_\SO$. As the network input, we then take the mean fluid
velocity, relative to the solids velocity in $\OO(\BO)$, i.e, 
\begin{equation*}
  \textbf{rel vel} \coloneqq \int_{\OO(\BO)}\ub - U^\text{total}\dif\x.
\end{equation*}
We have found $d_\OO = 4d_\BO$ to be an appropriate choice. For the input 
we then de-dimensionalise the data and take $\textbf{input} = \textbf{rel vel}
/ v$, where $v$ is the characteristic velocity, which we take as the
terminal settling velocity $v=0.047\si{\m\per\s}$.

Assuming that the mesh size is not
coarser than the size of the triangular rigid body, we can then safely apply
the isoparametric mapping technique \cite{Leh16, Leh17} to accurately integrate
over this domain in the CutFEM setting, without distorting the three level sets
describing the sides of the triangular body.

As the forces only depend on the angle of attack between the mean (integrated) 
flow relative to the triangles velocity and the orientation of the triangle, we
shall consider the input in a reference configuration. For this we choose the
bottom side of the triangle to be parallel to the $x$-axis. As a result,
the network learns the forces resulting from any angle of attack in this
reference orientation. To obtain the physical forces, we therefore rotate the
input into the reference configuration and rotate the drag/lift predictions
back into the physical orientation.
We also refer to \autoref{fig.domain-sketsches} for a sketch
of this configuration. Since the torque in two spatial dimensions is a scalar 
quantity, it is invariant with respect to rotation. 

\subsubsection{Network output}
\label{sec.network:subsec.design:subsubsec.output}
To keep the network general, we train the network to learn the dimensionless
drag, lift and torque coefficients
\begin{equation*}
  C_\text{drag} = \frac{2}{v^2\rho_\FL L}\bm{F}_1,
  \qquad
  C_\text{lift} = \frac{2}{v^2\rho_\FL L}\bm{F}_2
  \qquad\text{and}\qquad
  C_\text{torque} = \frac{4}{v^2\rho_\FL L^2}\bm{T}.
\end{equation*}
The reference speed is taken as the terminal velocity established below as
$v=0.047\si{\m\per\s}$ and the reference length $L=d_\SO=0.002\si{\m}$.

\subsubsection{Network architecture}
\label{sec.network:subsec.design:subsubsec.architecture}
\begin{figure}
  \centering
  \includegraphics[height=5.2cm]{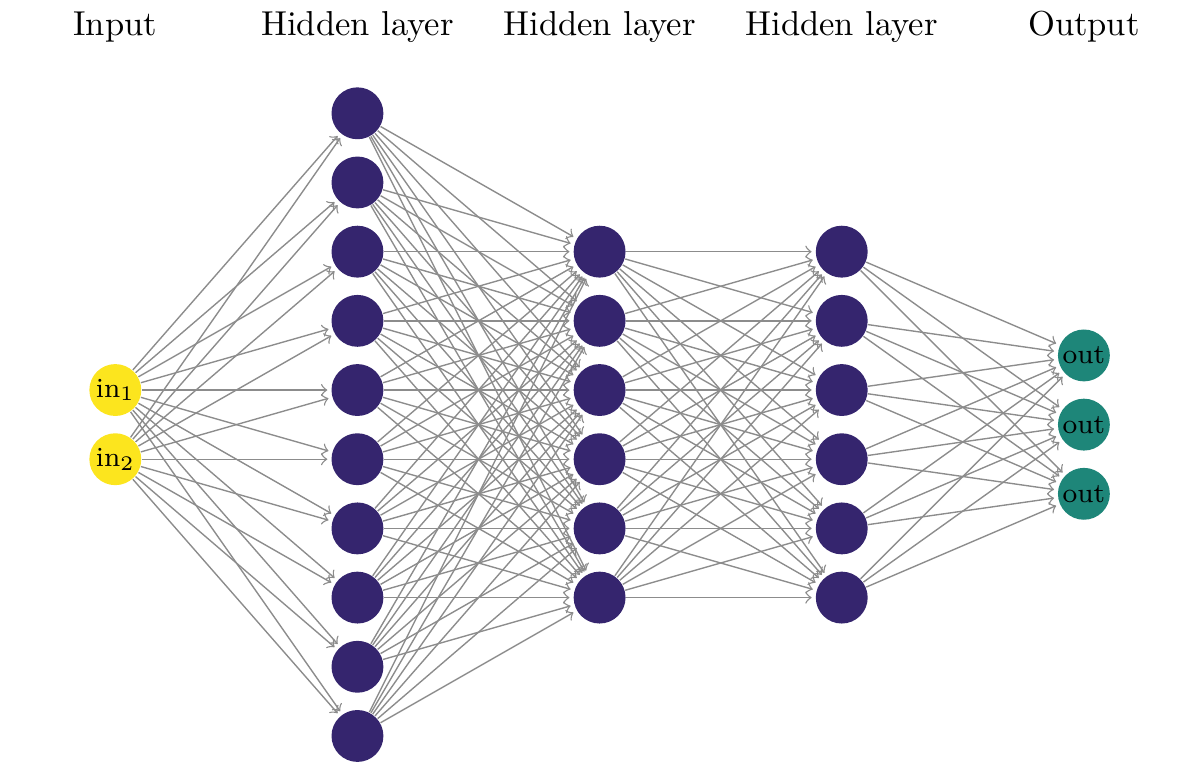}
  \caption{Illustration of a fully connected feed-forward network with two 
    inputs, three hidden layers consisting of 10, 6, and 6 neurones 
    respectively and a three outputs.}
  \label{fig.neural-network}
\end{figure}

We shall consider a fully connected feed-forward network with at least three
hidden layers and the ReLU activation function, i.e., $f(x)=\max\{0,x\}$.
An example of such a network can be seen in \autoref{fig.neural-network}. We
shall refer to networks by the number of neurones. For example, with $l_1/l_2/l_3$
we denote a network with three hidden layers consisting of $l_1$, $l_2$ and $l_3$
neurones respectively.
The number of neurones per layer and the number of layers we shall need for our
network will be determined experimentally, by inspecting the results achieved
during training. See \autoref{sec.network:sebsec:train} below.

\subsection{Training Data}
\label{sec.network:subsec.data}

In order to train a network, we need to generate the appropriate input and
target data.

\subsubsection{A priori computations}
\label{sec.network:subsec.data:subsubsec.apriori}

Neural networks are generally only accurate if they are applied in scenarios
which lie in the data range with which the network was trained. We therefore
need some information on how fast a triangular rigid body will fall under the
acceleration due to gravity as described in \autoref{sec.equations}  with the
material parameters described in \autoref{sec.equations:subsec.param}.

To get a better sense of the speeds with which we
need to generate the training data with, we consider an ALE
discretisation of a single triangular rigid body
falling without rotation in a viscous fluid. To this end we consider the domain
$\O=(0, 0.1)\times(0,0.2)$ and a triangular particle of side-length $0.002$
with the centroid $(0.05, 0.15)$ at $t=0$. For the fluid and solid parameters
we take the parameters described in \autoref{tab.parameters}. We then solve a
simplified
(restricted to vertical motion)
form of the system  \eqref{eqn.navier-stokes}, \eqref{eqn.solid-ode} and 
\eqref{eqn.navier-stokesBCs} until $t=10$.
To solve the coupled
fluid-solid problem, we consider a partitioned approach, i.e., we solve the
fluid and solid problem separately and iterate between the two until the
system is solved implicitly, see \autoref{sec.num} below and \cite{vWRFH21}.%
To make the ALE mapping simple,
we only allow motion in the vertical direction, i.e., we ignore the effects of
horizontal drag and torque. For the construction of such a mapping we refer to
\cite{vWRFH21}. We consider this set up for 10 different angles of
attack of the triangle and evaluate the terminal velocity.

For this computation we use a mesh with $h_{max}=\num{1e-3}$ in the bulk and
$h_{max} =\num{2.5e-4}$ on the triangle boundary with $\PP_5/\PP_3^{dc}$ finite
elements, resulting in a FE space with\footnote{\texttt{dof} denotes the
degrees of freedom of the full finite element space. 
$\texttt{dof}_\texttt{cond}$ denotes the number of degrees of
freedom remaining after static condensation of unknowns internal to each
element, i.e, the size of the system to be solved in every time-step.}
$\texttt{dof}=\num{211658}$ and $\texttt{dof}_\texttt{cond}=\num{84230}$. 
The time-derivative and mesh velocity are approximated using the BDF1 scheme
and the time-step used is $\dt=\nf{1}{500}$.

In the resulting computations, we observed that the maximal terminal velocity
of the triangle was $0.047\si{\m\per\s}$. This gives us a good indication for
the maximal velocity we need to consider to generate training data with.

\subsubsection{Generating training data}
\label{sec.network:subsec.data:subsubsec.generation}

In order to create a network of which we can expect sufficient accuracy, we
generate the training data in a setting which is as close to the application as
possible. However, in contrast to the following application of the
network, where as coarse as possible discretisations are
used for efficient simulations, training data is obtained on highly resolved 
meshes.

\begin{figure}
  \centering
  \includegraphics[width=4cm]{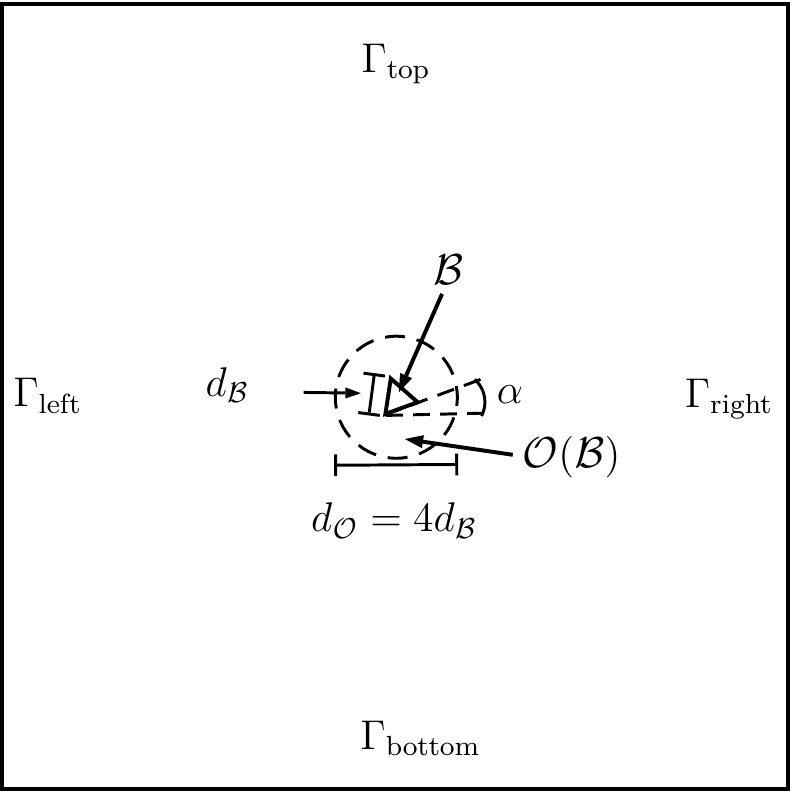}
  \caption{Sketch of the computational domain to generate the learning
    data set.}
  \label{fig.domain-sketsches}
\end{figure}

\paragraph{Set-up: Translational motion}
To generate learning data for translational motion, we take the
domain $(0, 0.5)^2$ with the equilateral triangular obstacle located at $(0.25,
0.25)$ in the reference configuration. This rigid body is then moved from $(0.1,
0.25)$ to $(0.4, 0.25)$ and back again over a time interval $[0, t_{end}]$. 
To get a wide range of relative velocities between the triangle and the mean
flow around the triangle, we accelerate the particle at different rates, i.e.,
consider different values for $t_{end}$. The physical location of the body is
then given by
\begin{equation*}
  (0.25 - 0.15\cos\big(\frac{\pi t}{2 t_{end}}\big), 0.25).
\end{equation*}
for $t \in [0, t_{end}]$. To implement this motion, we use a prescribed ALE
mapping, as above. 
In order to generate the data with different angles of attack, we rotate the
rigid body by an angle $\alpha$ around it's centre of mass
and rotate the resulting relative velocities and forces back into
the reference configuration. As a result, we can reuse the same ALE mapping to
simulate different directions of relative motion of the solid body.
A sketch of this configuration in the ALE reference configuration can be seen
in \autoref{fig.domain-sketsches}.

\paragraph{Set-up: Rotational motion}
Since the above ALE computations only include translational motion, this is
equivalent to only considering a parallel flow around a fixed obstacle. For
the network to be universally usable, we therefore also need to include
rotational flow data. This corresponds to rotating the triangle.
To this end, we consider a triangular rigid body located at the centre of the
domain $\O=(0,0.1)^2$. This is then rotated at different speeds, clockwise and
anti-clockwise, such that the total rotation at time $t$ with respect to the
initial configuration is given by $\sin(2\pi\cdot t / t_{end})$.

In this situation, using ALE is not suitable, as relatively small rotations
will lead to mesh-entanglement. We therefore use a highly resolved moving
domain CutFEM simulation, to generate data with rotational input. Further
details of this method are given below in \autoref{sec.num}.

\paragraph{Validation}
In order to ensure that the generated learning data is computed sufficiently
accurate, we consider the above cases over a series of meshes and
time-steps. For this test case, we take $t_{end}=2.0$ for both set-ups and
$\alpha=0$ for the translational set-up.

For the ALE discretisation, we take $\THk{4}$ elements on a mesh
with diameter $h_{max}$ in the bulk and a local mesh parameter $h_{loc}$ on the
boundary of the rigid body. In time we discretise using the BDF1 scheme with
the time-step $\dt$.

For the CutFEM discretisation, we take $\THk{2}$ elements on a mesh with global
meshing parameter $h_{max}$ and performing 3 levels of mesh bisections in the
domain where we compute the average relative velocity with an additional 5
levels of mesh bisections in the bounding circle of the rotating triangle.

The results for the convergence study can be seen in
\autoref{tab.data.validation}. For the ALE computations, we see that the
discretisation is accurate and that the second mesh already provides 2 -- 3
significant figures of accuracy in the target data. Since the neural network
prediction will introduce an additional approximation error, we consider this
to be sufficiently accurate for the training data. For the rotational CutFEM
data, we see that the forces are (in absolute value) significantly smaller than
the translational data. However we also see, that the second finest
discretisation should be accurate enough for the training of the network.

\begin{table}
  \centering
  \sisetup{
  round-precision=4,
  table-format=1.4e2,
  }
  \begin{tabular}{c lll ccc}
    \toprule
    Case(Method) & $h_{max}$ & $h_{loc}$ & $\dt$ & $C_{\text{drag},\text{max}}$
    &  $C_{\text{lift},\text{min}}$ & $C_{\text{torque},\text{max}}$\\
    \midrule
    Translational (ALE) 
    & 0.02 & $2\cdot10^{-4}$ & $\nf{1}{250}$ &
      \num{2.6094e+02} & \num{-5.9548e+01} & \num{2.9674e+01}\\
    & 0.01 & $1\cdot10^{-4}$ & $\nf{1}{500}$ &
      \num{2.6105e+02} & \num{-5.9673e+01} & \num{2.9699e+01}\\
    & 0.005 & $5\cdot10^{-5}$ & $\nf{1}{1000}$ &
      \num{2.6111e+02} & \num{-5.9751e+01} & \num{2.9694e+01}\\
    \cmidrule(lr){1-7}
    Rotational: (CutFEM)
    & 0.0084 & -- & $\nf{1}{250}$ &
      \num{1.6025e-02} & \num{-1.3880e-02} & \num{3.3697e-01}\\
    & 0.0042 & -- & $\nf{1}{500}$ &
      \num{1.3443e-02} & \num{-1.5876e-02} & \num{3.5969e-01}\\
    & 0.0021 & -- & $\nf{1}{1000}$ &
      \num{1.0735e-02} & \num{-8.9824e-03} & \num{3.5636e-01}\\
    \bottomrule
  \end{tabular}
  \caption{Validation results of the discretisation to generate the training
    data set.}
  \label{tab.data.validation}
\end{table}

\begin{remark}
It is known, that in general, the (fully) implicit BDF1 method is not a good
choice to discretise the time-derivative in the Navier-Stokes equations, since
the scheme is too diffusive, thus preventing for example vortex shedding.
However, the choice of BDF1 here is not problematic, since we are considering
flows with small Reynolds numbers. Furthermore, we have tested computations
with the less diffusive BDF2 scheme and no significant differences could be
observed in the solution.
\end{remark}

\paragraph{Generation}
To generate the data set, we consider $t_{end}\in\{2, 2.5, 3, 4, 6, 8, 1\}$
and rotate the triangle with angle $\alpha\in\{\frac{2i\pi}{3\cdot40}\st
i=0,\dots,39 \}$. Since the triangle is equilateral, the remaining angles of
attack $\alpha\in[\nf{2\pi}{3}, 2\pi)$ can be obtained by post-processing the
data appropriately. 
Based on the above validation computations, we consider the
mesh with $h_{max}=0.01$ and take the time-step $\dt=\nf{1}{500}$.

The resulting data set then contains $\num{2130000}$ input/output pairs.

For the rotational data we choose $t_{end}\in\{ 0.5, 1, 2, 3, 5, 6\}$. 
Again, from the above computations, we take the mesh with $h_{max}=0.0042$ and
the time-step is chosen as $\dt=\nf{1}{500}$.
As a result, we then obtain an additional $\num{8792}$ data sets.

\subsection{Training}
\label{sec.network:sebsec:train}

We implement the neural network described in
\autoref{sec.network:subsec.design}
using \texttt{PyTorch} \cite{PGM+19}. We use the mean squared error as the loss
function and take the Adam algorithm as the optimiser with a step size of
$10^{-4}$. The network is trained for a total of 20000 epochs. For the network 
to be able to predict all three values at the same time, we scale both the 
input and output data to be in the interval $[-1,1]$. 
In practice, the predictions are then scaled back appropriately, so that the
appropriate coefficients are obtained.

To ensure that the we do not over-fit the network to the training data set, we
additionally generated a validation data set in the same fashion as the
translational part of the
training data set but with different angles of attack and values for $t_{end}$
This then consists of $\num{431766}$ data points. During training we then also
evaluate the network on the validation data set and ensure that the the
training error does not decrease while the validation error increases.

The errors of the predictions made by the networks on the training and
validation data sets can be seen in \autoref{tab.training-error:single-nets}
for the case where a single network was used to predict drag lift and torque,
while \autoref{tab.training-error:separarte-nets} shows the error for separate
networks for each force. To make the results on data sets of different sizes
comparable, we use the norm $\nrm{\text{err}}{mean}^2 \coloneqq
\frac{1}{N}\sum_{i=1}^{N}(\text{prediction}_i-\text{value}_i)^2$ 
and $\nrm{\text{err}}{\infty} = \max_{i=1,\dots,N}(\vert
\text{prediction}_i-\text{value}_i \vert)$.

To find an appropriate network size, which is large enough to capture all the
information contained in the data set, while also being small enough for fast
evaluations in the final solver, we consider a number of different networks.
The chosen number of layers and neurones per layer can be seen in the first
column of \autoref{tab.training-error:single-nets}.

\begin{table}
  \centering
  \begin{tabular}{ccl SSSS}
    \toprule
    &&& \multicolumn{2}{c}{Training data} & 
      \multicolumn{2}{c}{Validation data}\\
    \cmidrule(lr){4-5} \cmidrule(lr){6-7}
    Architecture & Parameters & \multicolumn{1}{c}{Target} & $\nrm{\cdot}{mean}$
      & $\nrm{\cdot}{\infty}$ & $\nrm{\cdot}{mean}$ & $\nrm{\cdot}{\infty}$ \\
    \midrule
    30/20/10 & 953 & $C_\text{drag}$ & 1.658e+00 & 1.726e+01 & 1.867e+00 &
      1.191e+01\\
      & & $C_\text{lift}$ & 1.640e+00 & 1.883e+01 & 1.773e+00 & 1.460e+01\\
      & & $C_\text{torque}$ & 5.872e-01 & 5.006e+00 & 6.573e-01 & 4.293e+00\\
    \cmidrule(lr){4-7}
    50/20/20 & 1653 & $C_\text{drag}$ & 1.047e+00 & 8.618e+00 & 1.186e+00 &
      7.876e+00\\
      & & $C_\text{lift}$ & 1.048e+00 & 8.918e+00 & 1.174e+00 & 8.123e+00\\
      & & $C_\text{torque}$ & 5.560e-01 & 4.160e+00 & 6.253e-01 & 3.746e+00\\
    \cmidrule(lr){4-7}
    50/20/20/10 & 1833 & $C_\text{drag}$ & 1.186e+00 & 1.063e+01 & 1.324e+00 
        & 7.916e+00\\
      & & $C_\text{lift}$ & 1.161e+00 & 8.617e+00 & 1.316e+00 & 7.875e+00\\
      & & $C_\text{torque}$ & 5.613e-01 & 4.501e+00 & 6.320e-01 & 3.758e+00\\
    \cmidrule(lr){4-7}
    100/50/20/20 & 6853 & $C_\text{drag}$ & 9.584e-01 & 8.751e+00 & 1.080e+00 
        & 8.022e+00\\
      & & $C_\text{lift}$ & 9.681e-01 & 8.602e+00 & 1.088e+00 & 7.879e+00\\
      & & $C_\text{torque}$ & 5.511e-01 & 4.085e+00 & 6.201e-01 & 3.757e+00\\
    \cmidrule(lr){4-7}
    100/50/50/20 & 8983 & $C_\text{drag}$ & 9.726e-01 & 8.552e+00 & 1.098e+00 
        & 7.798e+00\\
      & & $C_\text{lift}$ & 9.582e-01 & 8.765e+00 & 1.074e+00 & 8.006e+00\\
      & & $C_\text{torque}$ & 5.507e-01 & 3.963e+00 & 6.194e-01 & 3.577e+00\\
    \bottomrule
  \end{tabular} 
  \caption{Prediction errors in a weighted $\ell^2$ norm and the maximum norm 
    on the training and validation data sets after 20000 epochs of training 
    for a number of different network sizes. Each network predicts all three
    functional values simultaneously.}
  \label{tab.training-error:single-nets}
\end{table}

\begin{table}
  \centering
  \begin{tabular}{l SSSS}
    \toprule
    & \multicolumn{2}{c}{Training data} & \multicolumn{2}{c}{Validation data}\\
    \cmidrule(lr){2-3} \cmidrule(lr){4-5}
    \multicolumn{1}{c}{Target} & $\nrm{\cdot}{mean}$ & $\nrm{\cdot}{\infty}$ & 
      $\nrm{\cdot}{mean}$ & $\nrm{\cdot}{\infty}$ \\
    \midrule
    $C_\text{drag}$ & 9.591e-01 & 8.832e+00 & 1.082e+00 & 8.095e+00\\
    $C_\text{lift}$ & 9.959e-01 & 8.672e+00 & 1.124e+00 & 7.868e+00\\
    $C_\text{torque}$ & 5.538e-01 & 4.143e+00 & 6.233e-01 & 3.631e+00\\
    \bottomrule
  \end{tabular}
  \caption{Prediction errors in a weighted $\ell^2$ norm and the maximum norm 
    on the training and validation data sets after 20000 epochs. Separate networks
    of the architecture 50/20/20 with 1611 parameters are used to predict the
    three functional values.}
  \label{tab.training-error:separarte-nets}
\end{table}

Looking at the results in \autoref{tab.training-error:single-nets}, we see that
the these are broadly similar for all six considered network architectures.
For the three layer networks, we observe that the prediction error
resulting from the 30/20/10 network are almost 2 times larger than that
of the 50/20/20 network.
Looking at the four layer networks, we see that the errors are generally the
same as for the 50/20/20 network, with some small deviations in both
directions.

In order to check whether we can gain more accuracy by considering separate
networks for the three functionals, we train three separate 50/20/20 networks.
The results thereof can be seen in \autoref{tab.training-error:separarte-nets},
where we observe that the prediction errors are about the same as those
realised by the single network with the same architecture in
\autoref{tab.training-error:single-nets}.

\begin{figure}
  \centering
  \includegraphics{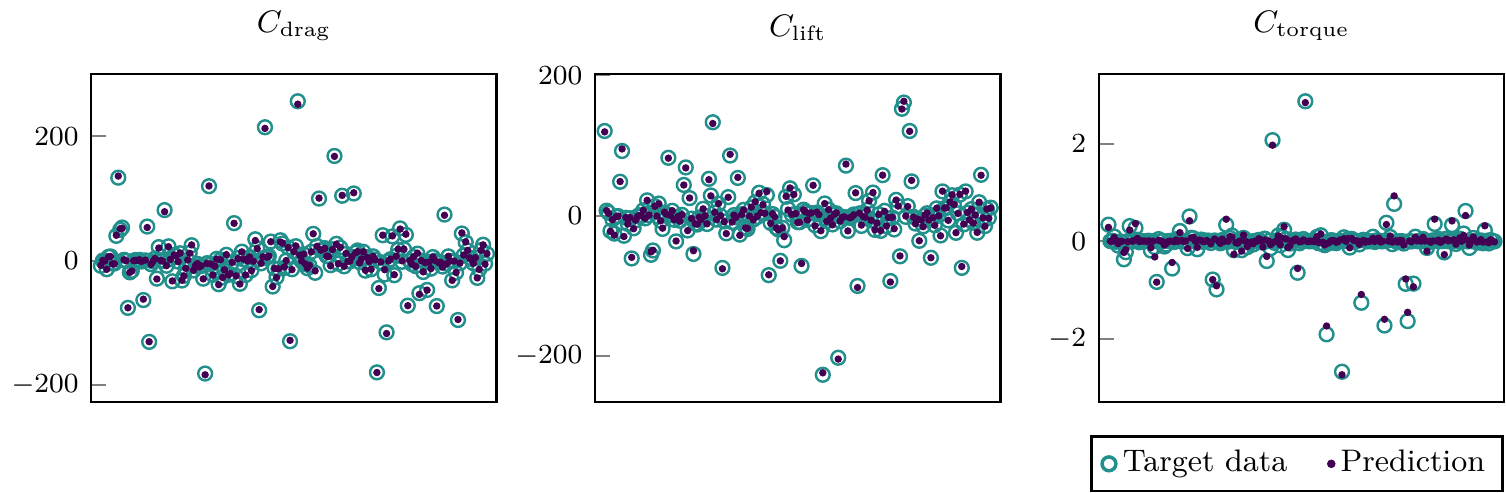}
  \caption{Target data and network prediction for 300 random points in the
    training data set. Network architecture: 50/20/20, 1653 parameters.}
  \label{fig.prediction}
\end{figure}

Both in \autoref{tab.training-error:single-nets} and
\autoref{tab.training-error:separarte-nets} we observe that the errors are
generally similar and very large.
To give these errors more
meaning, we plot the predictions and the target values for a random selection
of point from the validation data set in \autoref{fig.prediction}. Here we see
that in general the predicted values match the target values well. 
This indicates that the size of the errors in
\autoref{tab.training-error:single-nets}
and \autoref{tab.training-error:separarte-nets} are due to the size of the
target values. The errors in \autoref{tab.training-error:single-nets}
indicate that overall the force dynamics are captured with about 1 -- 2
significant figures of accuracy.

\begin{figure}
  \centering
  \includegraphics{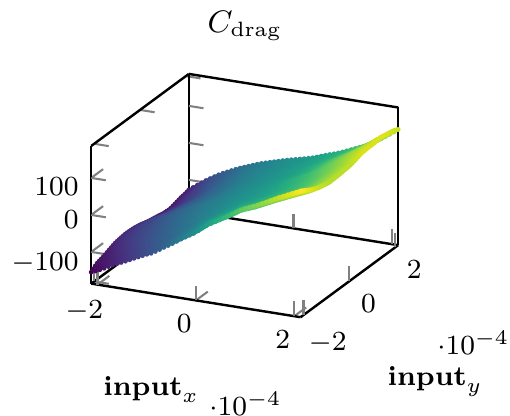}
  \includegraphics{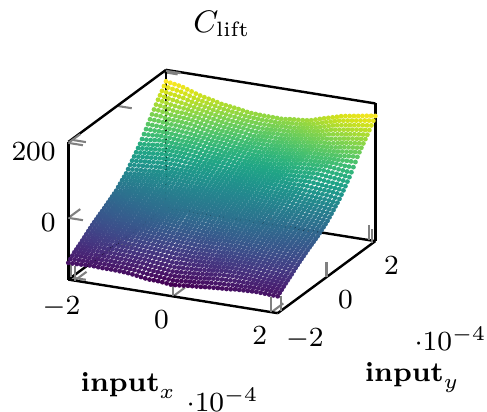}
  \includegraphics{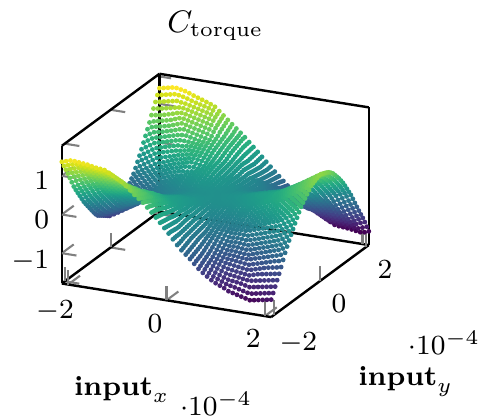}
  \caption{Network prediction components from the 50/20/20 network
   as functions of input variables.}
  \label{fig.network.as.function}
\end{figure}

As a result of the above considerations, we choose the single 50/20/20
network for our application of predicting the drag lift and torque from the
average velocity around the rigid particle. 
Considering this network as a
function $\R^2\rightarrow\R^3$, we can then plot the individual function
components as a function of the input in a three dimensional plot. This can be
seen in \autoref{fig.network.as.function}. This illustrates that while the drag
and lift coefficients are represented by a relatively simple function, the
torque coefficient is significantly more involved.

Due to the simple input for the
network, requiring two integrals over a relatively small area and a
rotation of these two values into the reference configuration, as well as the
small size of the network itself, the additional computational effort introduced
to the solver by predicting the forces, rather than evaluating them directly,
is negligible compared to the effort required to solve the non-linear system
resulting from the FEM discretisation of the Navier-Stokes equations.

\begin{remark}[Training times]
The training the above networks was performed on a Tesla V100 PCIe 16GB
graphics card with \texttt{PyTorch} using \texttt{CUDA} version  10.1. Due to
the small size and simple structure of the networks, the training times for
20000 epochs ranged between 277 seconds for the 30/20/10 network and 936
seconds for the 100/50/20/20 network.
\end{remark}

\subsection{Validation}
\label{sec.network:subsec.eval}

In the previous section we have compared the network predictions against the 
``true'' values in the sense of the ALE training data.
However, the aim for the network is to predict the force
values in a CutFEM simulation and to be more accurate than the boundary
integral evaluation. 

To validate that the neural network predictions are in fact more accurate than
the direct evaluation of the forces from the boundary integral,
we run a moving
domain CutFEM simulation of the set up, with which we generated the training
data for $t_{end}=3$. Here we took a background mesh with $h_{max}=10^{-3}$ in
the area
where we compute the average of the velocity around the triangle and
$h_{max}=0.04$ in the remaining part of the domain. The mesh in the averaging
area is therefore a factor of 2 smaller than the size of the rigid body. 
On this mesh we work with unfitted $\THk{2}$ and $\THk{3}$ elements.
We then compute the errors of the force prediction and evaluation, by
comparing the values against the direct evaluation of a highly resolved ALE
computation of the identical set-up.
The spatial discretisation is as for the training data generation above.
In both cases, the time-step $\dt=\nf{1}{300}$ was chosen.

\begin{figure}
  \centering
  \includegraphics{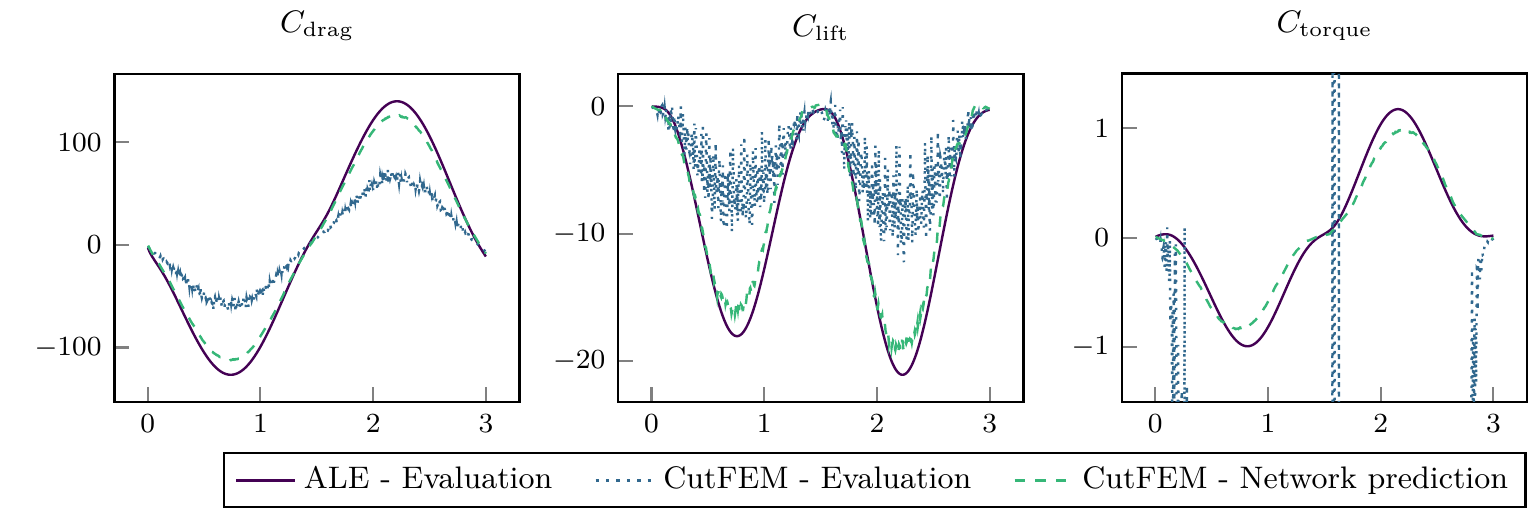}
  \caption{Prediction and evaluation in a CutFEM simulation with $\THk{2}$
    elements compared against the values evaluated in a fitted ALE 
    computation.}
  \label{fig.predction-error-CutFEM}
\end{figure}

The resulting prediction errors of the forces in the CutFEM simulation
are given in \autoref{tab.prediction-error-CutFEM}. 
Here we find that mean and maximal prediction error are at least one order of 
magnitude smaller than the direct evaluation.
In \autoref{fig.predction-error-CutFEM} we plot the resulting forces for
a single run of the above computations ($\alpha=0$). This shows, that while
there is a visible difference between the prediction and the ``ground truth'',
the predictions mirror the real forces much better than the direct
force computation
via the boundary integral evaluation.
Here we also note, that there was no significant improvement in the direct
evaluation when using higher order elements.
We conclude that while the predictions are not perfect, we have significantly
improved on the direct computation of the forces 
on an under resolved computational mesh. 
However, we also note that we cannot expect any asymptotic mesh convergence
here, as the prediction error will begin to dominate once the interface is
sufficiently resolved in every time-step.

\begin{table}
  \centering
\begin{tabular}{cc SSSSSS}
    \toprule
    && \multicolumn{2}{c}{$C_\text{drag}$} &
      \multicolumn{2}{c}{$C_\text{lift}$} & 
      \multicolumn{2}{c}{$C_\text{torque}$}\\
    \cmidrule(lr){3-4} \cmidrule(lr){5-6} \cmidrule(lr){7-8}
    Force comp. & Discr. &  $\nrm{\cdot}{mean}$ & $\nrm{\cdot}{\infty}$ &
      $\nrm{\cdot}{mean}$ & $\nrm{\cdot}{\infty}$ & $\nrm{\cdot}{mean}$ &
      $\nrm{\cdot}{\infty}$\\
    \midrule
    Evaluation & $\THk{2}$ & 7.859e+00 & 8.408e+01 & 1.329e+00 & 3.321e+01 
      & 1.440e+02 & 2.285e+03\\
    Prediction & $\THk{2}$ & 1.543e+00 & 2.320e+01 & 2.264e-01 & 4.812e+00 
      & 1.478e-01 & 2.917e+00\\
    Evaluation & $\THk{3}$ & 6.589e+00 & 7.172e+01 & 1.471e+00 & 4.147e+01 
      & 2.411e+02 & 3.919e+03\\
    Prediction & $\THk{3}$ & 1.515e+00 & 2.279e+01 & 2.366e-01 & 5.050e+00 
      & 1.479e-01 & 2.902e+00\\
    \bottomrule
  \end{tabular}
\caption{Absolute errors in the forces resulting from the direct evaluation 
of the boundary integrals and the prediction made by the neural network in a 
CutFEM computation.}
  \label{tab.prediction-error-CutFEM}
\end{table}

\section{Numerical examples}
\label{sec.num}

We consider a number of numerical examples which use the neural network trained
in the previous section in an under resolved, moving domain, 
CutFEM discretisation.
The details and analysis of this method as applied to the
time-dependent Stokes equations on moving domains is given in \cite{vWRL20}.
The idea of this method is to use ghost-penalty stabilisation to implement a
discrete, implicit extension of the solution into the exterior of the fluid
domain, in order to apply a BDF formula to the time derivative in the
case of a moving interface.
The only modification here is the addition of the convective term. See also
\cite{vWRFH21}, where this method has also been applied to a fluid-structure
interaction problem with contact and in which the fluid is described by the
incompressible Navier-Stokes equations.

To solve the coupled fluid-solid system, we use a partitioned approach.
In the sub-step to update the solid velocity we use an update scheme with
relaxation, together with Aitken's $\Delta^2$ method~\cite{IT69} to determine
the appropriate relaxation parameter and thereby accelerate the relaxation
scheme.
A relaxation is necessary for stability of the partitioned solver
of the coupled system and different relaxation strategies, including
Aitken's $\Delta^2$ method are discussed in the literature~\cite{KW08}.

\begin{remnum}[Geometry description in CutFEM]
In CutFEM, numerical integration on elements which are cut by the level-set
function describing the geometry, is based on a piecewise linear interpolation
of the (smooth) level set function. This is necessary to robustly construct 
quadrature rules on cut elements, see \cite[Section~5.1]{BCH+14}. To represent
the three sides of the triangular rigid bodies, we therefore use three separate
level set functions $\phi_1$, $\phi_2$ and $\phi_3$.
\texttt{ngsxfem}~\cite{ngsxfem}
then provides the functionality to integrate with respect to each of these
level sets, for example the domain where all level sets are negative. Since the
level sets describing the sides are linear. the numerical integration on the
geometry posed here is exact and we do not have to construct a single level set
function $\phi=\max(\phi_1,\phi_2,\phi_3)$ which would then be interpolated
into the $\PP^1$ space on the mesh, which in turn would remove the sharp
corners of the particles.
\end{remnum}

\subsection{Example 1: Free-fall restricted to vertical motion}
\label{sec.num:subsec.ex1}
In \autoref{sec.network:subsec.eval} we validated the neural network approach
in the setting of prescribed solid motion.
To validate the method in our target setting of free-fall, we shall consider
the settling of a single particle, restricted to vertical motion as in
\autoref{sec.network:subsec.data:subsubsec.apriori}. We consider this
simplification, in order to compare the results against a resolved ALE
simulation.

\begin{figure}
  \centering
  \includegraphics[width=5cm]{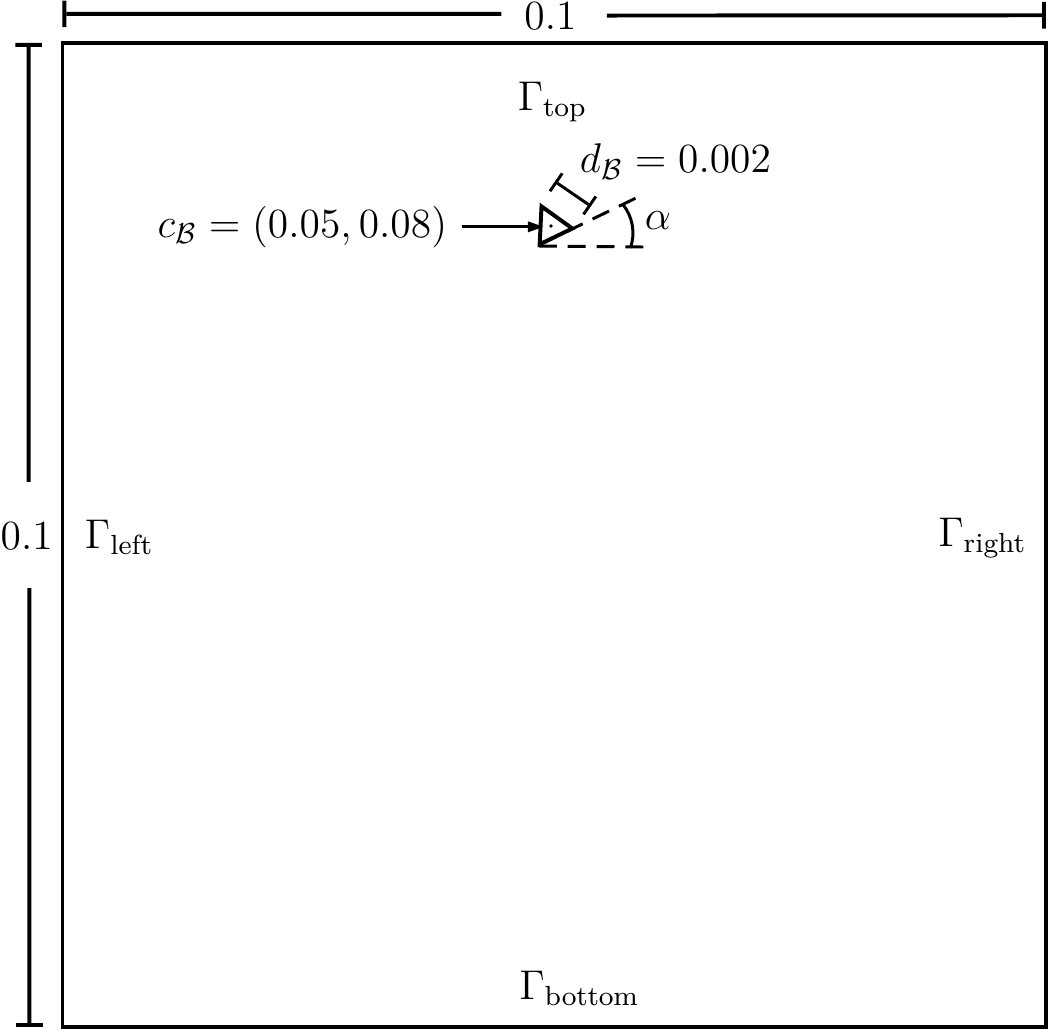}
  \hspace{2cm}
  \includegraphics[width=5cm]{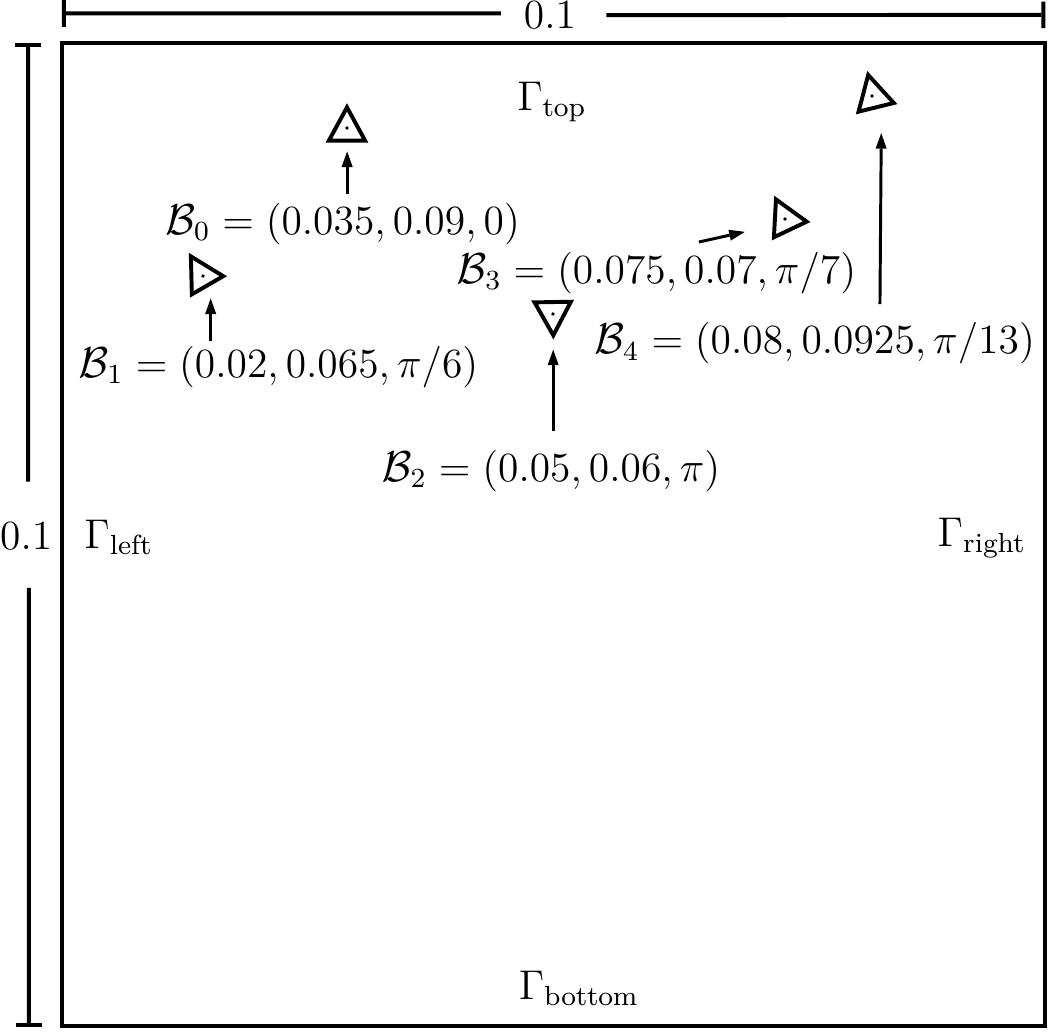}
  \caption{Domain sketches for the initial configurations of the computational
    examples. Left: A single particle (\autoref{sec.num:subsec.ex1} and 
    \autoref{sec.num:subsec.ex2}).
    Right: Five particles (\autoref{sec.num:subsec.ex3} 3).}
  \label{fig.numex-domain-sketches}
\end{figure}

To this end we take the domain $\O=(0,0.1)^2$, assert no-slip boundary
conditions at the left and right wall and the do-nothing condition
$\mu_\FL\nabla\ub\n = p\n$
at the top and bottom, see also~\cite{HRT92} 
for details thereon. The rigid body is an equilateral triangle with side-length
$2\cdot10^{-3}$. The fluid and solid material parameters are as in
\autoref{tab.parameters}. At $t=0$ the centre of mass of the rigid body is at
$(0.05, 0.08)$ and we rotate the body by an angle $\alpha$ with respect to
reference configuration, in which the bottom of the triangle is parallel to the
$x$-axis. We shall consider $\alpha=0, \pi/12,\pi/6$. 
See also the left sketch in \autoref{fig.numex-domain-sketches} for an 
illustration of this initial configuration.

For the ALE computation we consider $\THk{5}$ elements on a mesh with
$h_{max}=\num{0.005}$, $h=\num{0.00125}$ in a horizontal strip of height
$\num{8e-3}$ around the rigid body and $h=\num{0.0004}$ on the interface of the
rigid body.
Based on our validation experiments in
\autoref{sec.network:subsec.data:subsubsec.generation}, this discretisation is
sufficiently accurate to serve as a reference here.

To ensure that the observed motion is due to our neural network approach, 
rather than of the discretisation itself, we
again consider two different CutFEM simulations. In the first we shall base
the solid motion on the forces predicted by the neural network, while in the
other the forces are computed by the direct evaluation of the FEM solution.
For both CutFEM computations we consider a mesh with
$h_{max}=\num{1e-3}$ using $\THk{2}$ elements.
For the unfitted discretisation used, we take the Nitsche parameter to enforce
the Dirichlet boundary conditions on the level set interfaces
as $\gamma_N=100$ while the stability and extension ghost-penalty parameters
are $\gamma_{gp,stab}=0.1$ and $\gamma_{gp, ext}=0.001$ respectively. For
details on these parameters we refer to \cite{vWRL20}.

We take the time-step $\dt=\nf{1}{250}$ and compute until $t=1.0$.
For the partitioned fluid-solid solver, we take the initial relaxation
parameter as $\omega=0.5$, allow a maximum of 10 sub-iterations per time
step and a tolerance of $1\%$ in the relative update.

The forces acting from the fluid onto the rigid body are evaluated using the
single 50/20/20 network for drag, lift and torque. As discussed above we use
an isoparametric approach to accurately integrate the
fluid velocity relative to the solid velocity in $\OO(\BO)$.

\begin{figure}
  \centering
  \includegraphics{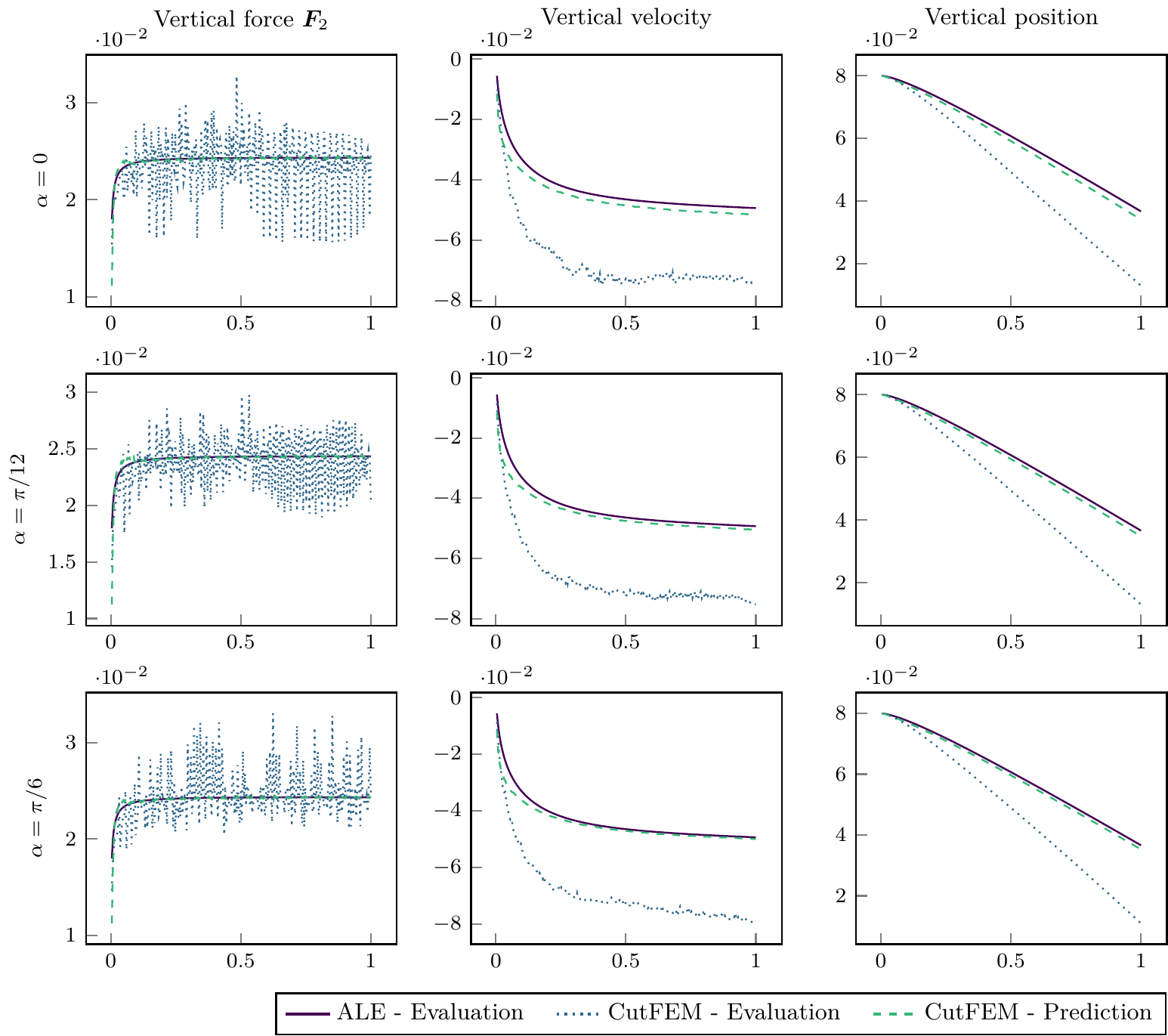}
  \caption{Vertical force, velocity and position of a single falling triangular
    rigid  body restricted to vertical motion. Comparison between an ALE
    reference computation, the force computation in an under resolved CutFEM
    computation (Evaluation) and the network prediction (Prediction)
    in an under resolved CutFEM 
    computation.}
  \label{fig:numex-ex1:vel-pos}
\end{figure}

In \autoref{fig:numex-ex1:vel-pos} we can see the vertical force, the
vertical
speed and the vertical position of the centre of mass of the rigid body over
time for each of the three triangle orientations. Here we see that for all
three orientations, the drag obtained by the network prediction is
significantly more accurate than the direct evaluation of the boundary
integral. Looking at the body's vertical speed and position, we see that there
is a visible difference between the ALE reference solution and the movement
resulting from the predicted forces. The uniformly faster speed in the CutFEM
prediction computations can be attributed to the fact, that at small speeds of
the particle, the network underestimated the drag, as can be seen in the left
column of \autoref{fig:numex-ex1:vel-pos}. However we also seen that the
predictions are significantly better than the result from the direct evaluation
of the forces. 
This clearly shows, that the neural network approach is able to realise
accurate results on very coarse and under resolved meshes, where the
unfitted approach with the standard evaluation of the resulting forces does not
lead to realistic motion.

\subsection{Example 2: Single triangular rigid body}
\label{sec.num:subsec.ex2}

We now consider the full fluid-rigid body system including translational
movement in all directions as well as rotational motion. 
As a result, we do not have an ALE reference to compare the results 
with. The initial configuration is therefore chosen as in
\autoref{sec.num:subsec.ex1}, with the initial rotation $\alpha=\pi/3$.

We again take the background mesh with $h_{max}=\num{1e-3}$ 
and second mesh with $h_{max}=7.5\cdot10^{-4}$
to investigate the mesh dependence. On each mesh we take $\THk{2}$ elements.
The remaining discretisation parameters are chosen as in the unfitted
simulation in \autoref{sec.num:subsec.ex1}.

\begin{figure}
  \centering
  \includegraphics{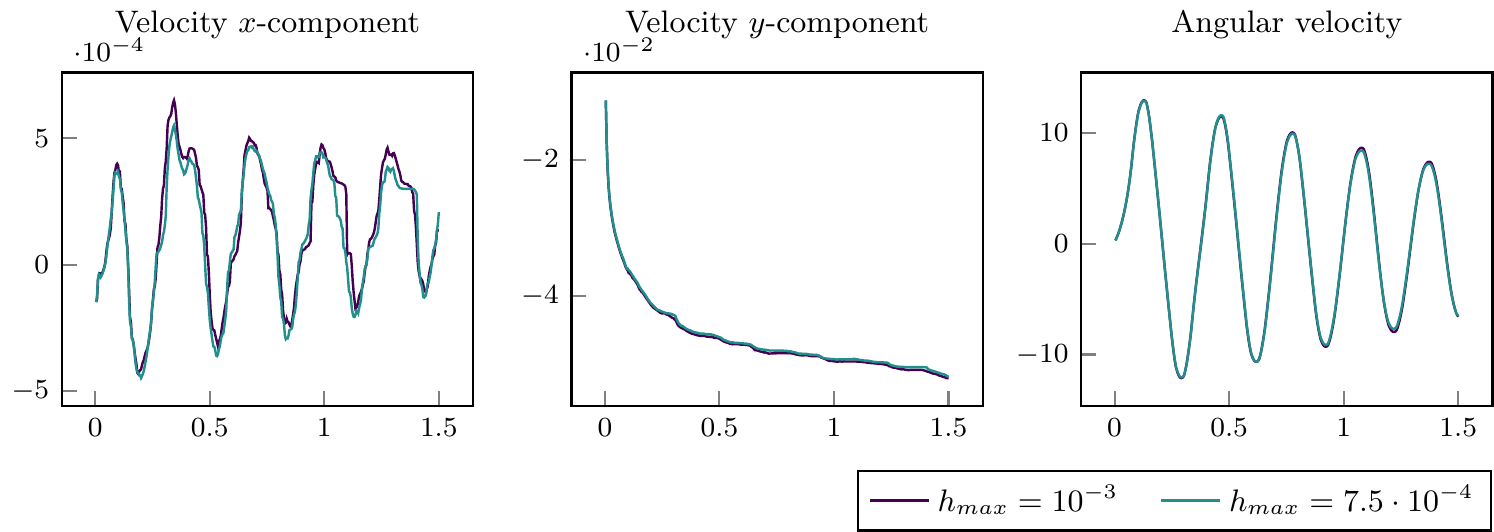}
  \caption{Translational and rotational velocity components of a single
    triangular rigid body in free fall on an under-resolved mesh with the
    forces from the predictions by a deep neural network.}
  \label{fig.numex2:solid-vels}
\end{figure}

\begin{figure}
  \centering
  \includegraphics[width=0.49\textwidth]{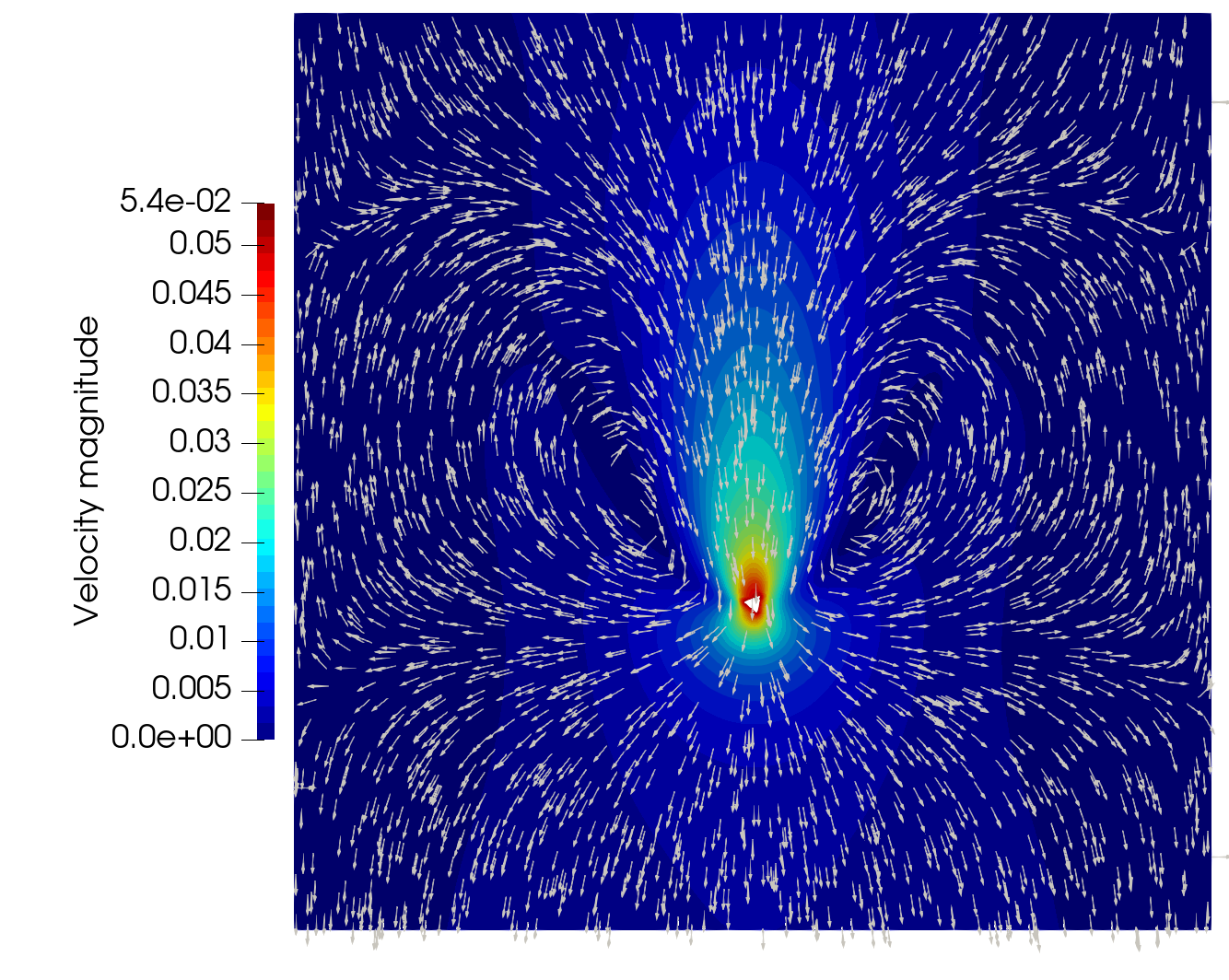}
  \hfill
  \includegraphics[width=0.49\textwidth]{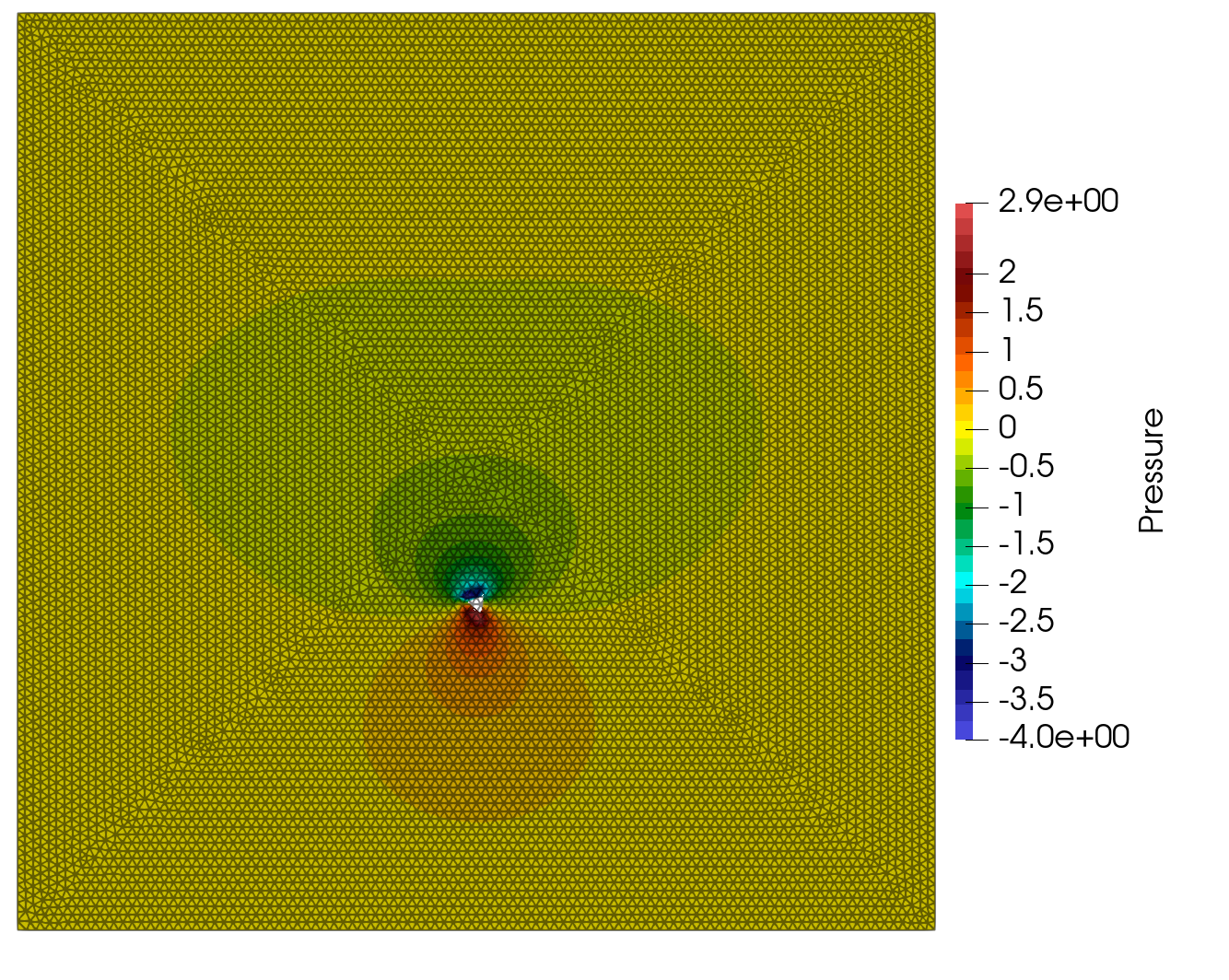}
  \caption{Velocity field (left) and pressure with mesh (right)
    at $t=1.0$ resulting from a single
    triangular rigid body in free fall on a mesh with $h_{max}=10^{-3}$ with
    the forces governing the solid motion obtained from a neural network.}
  \label{fig.numex2:fluid}
\end{figure}

In \autoref{fig.numex2:solid-vels} we see the resulting velocity components
which make up the movement of the rigid body for the two meshes considered. 
The fluid solution at time $t=1.0$ on the coarser of the two meshes is
visualised in \autoref{fig.numex2:fluid}.

Looking at velocity components in \autoref{fig.numex2:solid-vels}, we
immediately see that the influence of the mesh (in the considered under
resolved range) is very small. 
While this does not indicate accuracy of
the method, this does show that the method is stable.
As is to be expected, the vertical velocity
component dominates the translational motion. Furthermore, we see that while a
terminal velocity is not fully reached, that the the acceleration between
$t=0.5$ and $t=1.5$ is small and the velocity during this time is
very similar and only about 10\% faster compared to the a priori computations
is \autoref{sec.network:subsec.data:subsubsec.apriori} and in our validation
example with restricted motion in \autoref{sec.num:subsec.ex1}.

The angular velocity in \autoref{fig.numex2:solid-vels} appears very large at
first sight. Factoring in that the side-length of the body is $2\cdot10^{-3}$,
we find that the resulting maximal velocity at the corners of the triangular
rigid bodies is of order $10^{-2}$. So in general, the velocity of the triangle
resulting from rotation is smaller than the vertical velocity component. 

Overall, this example shows that our method leads to plausible results
on highly under resolved meshes, on which the standard CutFEM approach cannot
realise accurate forces and thereby cannot realise accurate solid motion.
Finally, we emphasise that a comparison to a fitted ALE simulation is out of
scope for this example, since this would have to include re-meshing procedures
and projections of the solution onto the resulting new meshes, in order to
avoid mesh-entanglement resulting from the rotation of the particles.

\subsection{Example 3: Multiple rigid bodies}
\label{sec.num:subsec.ex3}

As a final more advanced example, consider the same basic set-up as in 
\autoref{sec.num:subsec.ex2}, and take 5 rigid triangular rigid bodies denoted
by $\BO_i$, $i=0,\dots,4$.%
The particles initial position and orientation are then described 
by their centre of mass as the angle of rotation with
respect to the reference configuration, in which the bottom side of the
triangle is parallel to the $x$-axis. We denote the initial centre of mass
and rotation by $(c_{x}, c_y, \alpha)$.
We consider the initial states $(0.035, 0.09, 0)$, $(0.02, 0.065, \pi/6)$,
$(0.05, 0.06, \pi)$,  $(0.075, 0.07, \pi / 7)$ and
$(0.08, 0.0925, \pi / 13)$ for $\BO_0,\dots,\BO_4$ respectively.
A sketch of this configuration can be seen in the right of
\autoref{fig.numex-domain-sketches}.
The discretisation parameters of the moving domain CutFEM method are all as in
\autoref{sec.num:subsec.ex2}.

\begin{figure}
  \centering
  \includegraphics{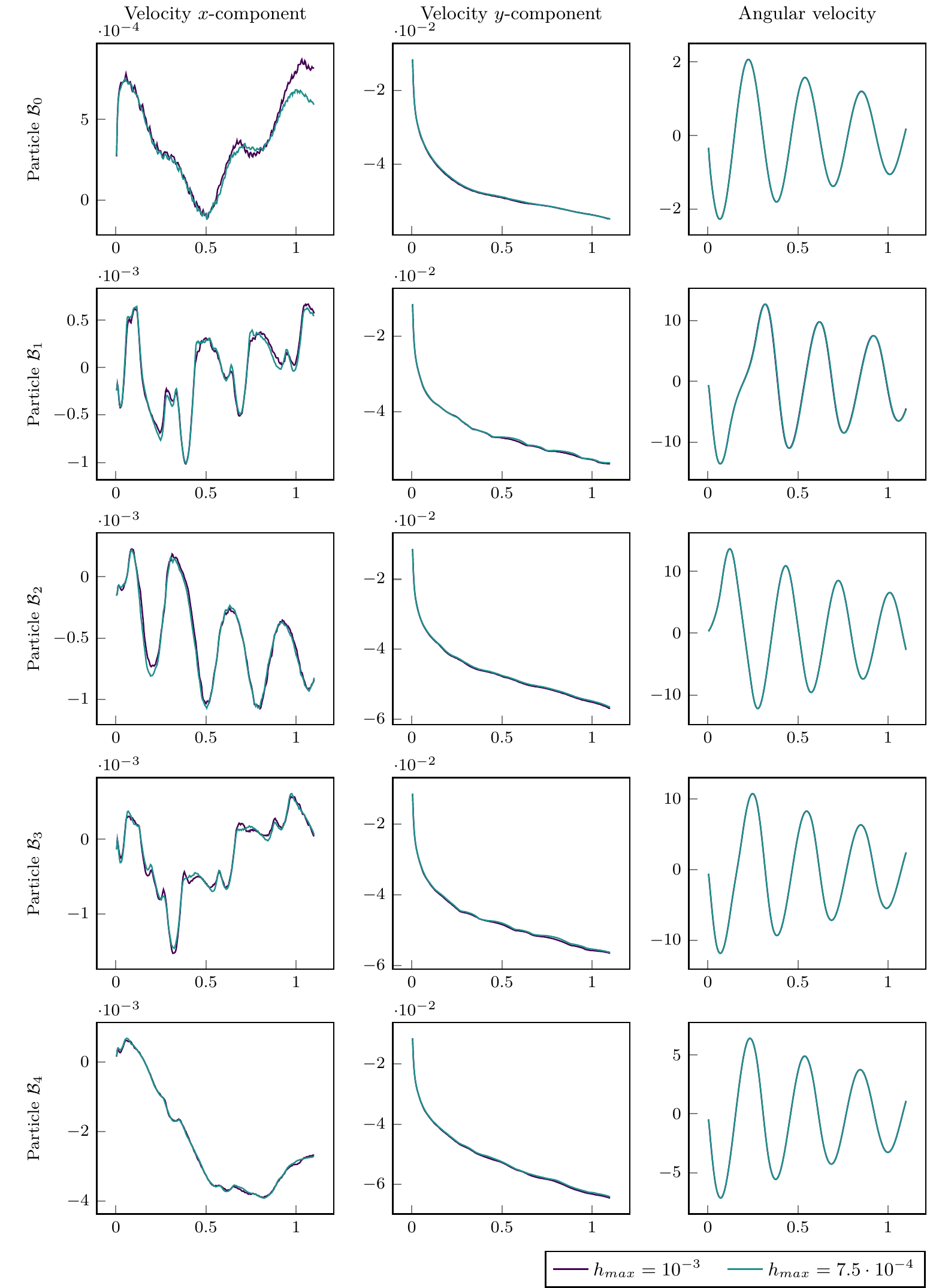}
  \caption{Translational and rotational velocity of five
    triangular rigid bodies in free fall in an under-resolved CutFEM
    simulation where the forces governing the solid motion are from the
    evaluation of a deep neural network.}
  \label{fig.numex3:solid-vels}
\end{figure}

\begin{figure}
  \centering
  \includegraphics[width=0.49\textwidth]{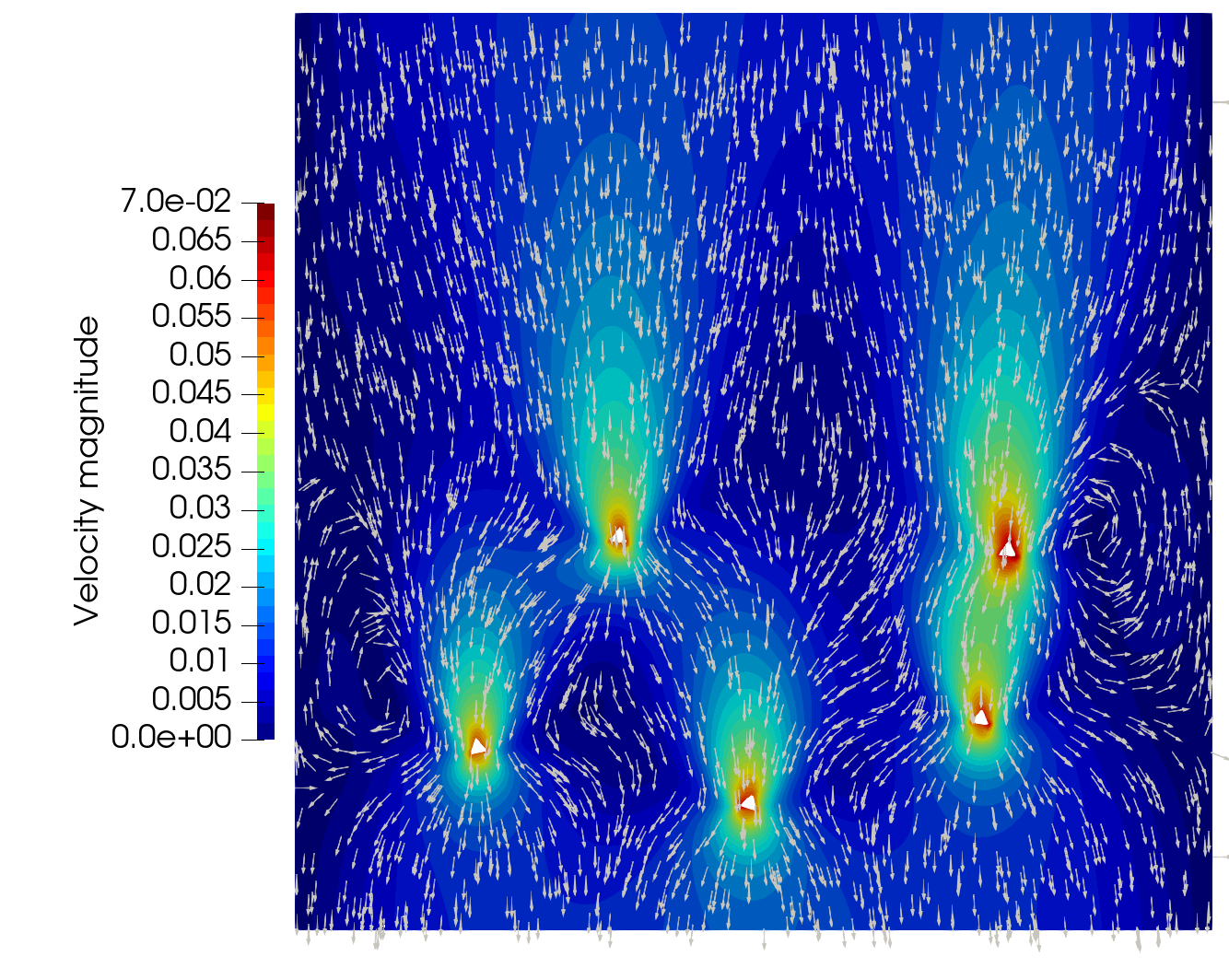}
  \hfill
  \includegraphics[width=0.49\textwidth]{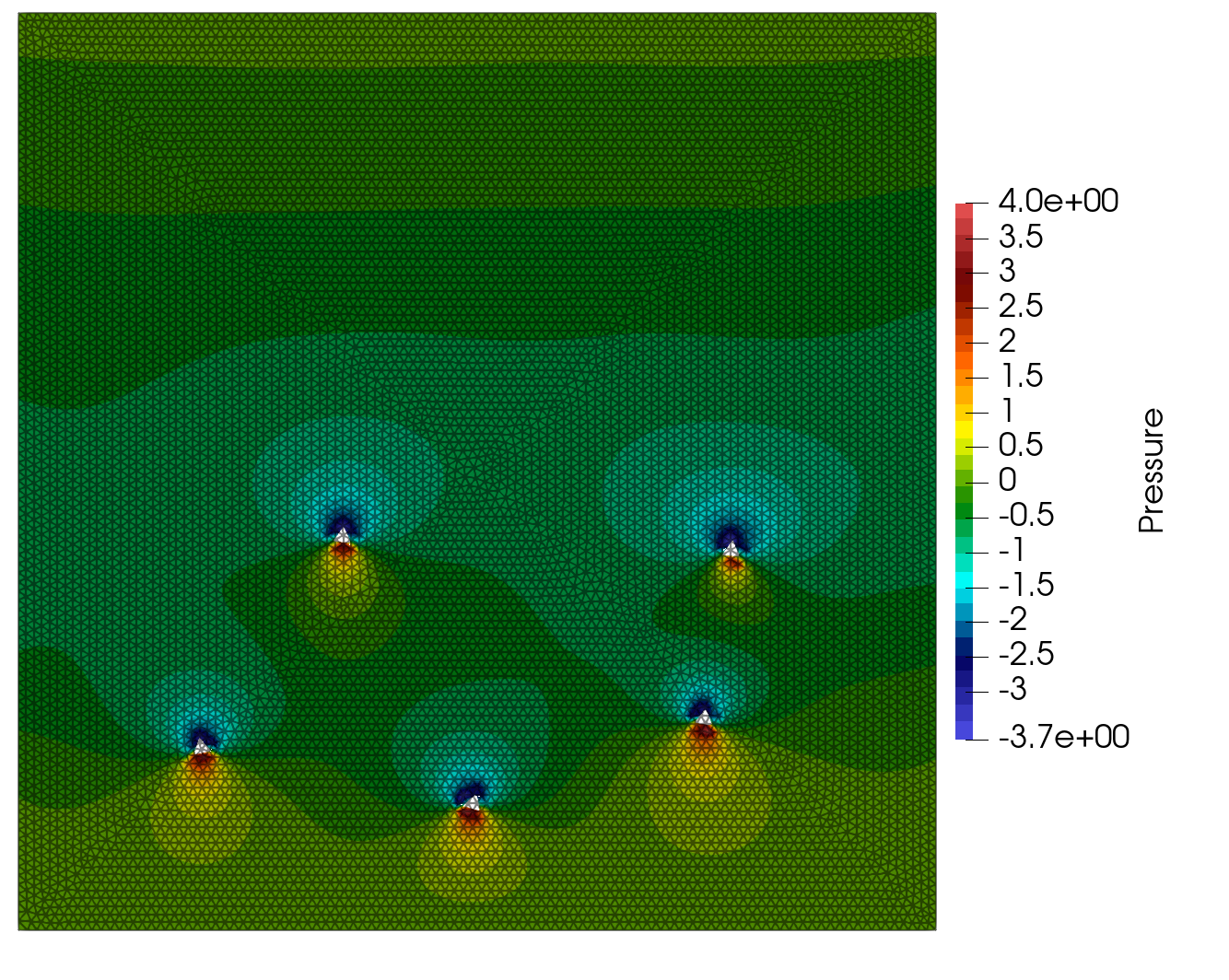}
  \caption{Velocity field (left) and pressure with mesh (right)
    at $t=1.0$ resulting from five
    triangular rigid bodies in free fall on a mesh with $h_{max}=10^{-3}$ with
    the forces governing the solid motion obtained from a neural network.}
  \label{fig.numex3:fluid}
\end{figure}

The results from these computations are shown in
\autoref{fig.numex3:solid-vels} and \autoref{fig.numex3:fluid}. The
translational and rotational velocity components are shown in
\autoref{fig.numex3:solid-vels} while the fluid solution at $t=1.0$ is shown in
\autoref{fig.numex3:fluid}.

In \autoref{fig.numex3:solid-vels} we see that there is again no large
dependence on
the mesh and that in general the translational
velocity of all particles are larger than in the single particle case above.
Nevertheless, the order of magnitude is the same as before. 
The slightly faster translational speeds 
of the particles
make sense, as each particle sets the fluid in motion which then helps to
transport the other particles. Furthermore we see both in
\autoref{fig.numex3:solid-vels} and \autoref{fig.numex3:fluid}, that $\BO_4$
has the largest (vertical) velocity. This observation is also
meaningful, as $\BO_4$ 
is directly in the wake of $\BO_3$. Furthermore, we note that the angular
velocity of $\BO_0$ and $\BO_4$, which are above the other particles, is
smaller than that of the other particles. We attribute this to the fact, that
these two particles are in the parallel flow wake of the other three particles,
resulting in a smaller torque acting on these two particles.

\section{Conclusions and Outlook}
\label{sec.conclusion}

\paragraph{Conclusions}
We have presented a framework using a hybrid finite element / neural network
approach to obtain improved drag lift and torque values acting on a triangular
rigid body. For this, we trained a small deep neural network using training
data obtained from prototypical motion configurations in resolved finite
element simulations. We considered a range of networks by varying the number
of layers as well as the number of neurones per layer. Here we saw that
the mapping from the average velocity around the particle to the forces acing
on this particle was accurately captured by relatively small networks.
Applying one of these networks in an under resolved moving domain
CutFEM simulation, we showed that this approach results in significantly
more accurate force values 
compared with the direct evaluation of the forces from the 
finite element solution.
We saw that the predicted values were on average
an order of magnitude more accurate than the direct evaluation of the forces
from the Navier-Stokes solution.

\paragraph{Outlook}
A number of interesting questions remain for future research. For example, it
would be very interesting to compare the particle motion realised in
\autoref{sec.num:subsec.ex2} and \autoref{sec.num:subsec.ex3} to fitted ALE
simulations using re-meshing techniques.

Furthermore, it remains to generalise this approach to other particle shapes
and sizes. Here it would also be interesting to see, if a different choice of
input data could generalise the approach to use a single neural network to
predict accurate forces with different material parameters.

Finally, the method should be extended to the interaction of several
closely neighbouring particles as well as to the particle-wall interaction. In
particular, extended training data sets have to be created for this purpose.
Also, it has to be explored whether it is sufficient to consider the mean
ambient velocity as network input in this case or whether further information,
e.g. the velocity gradient or acceleration rates of the particles or
the fluid should be taken into account.

\section*{Acknowledgements}
The authors acknowledge support by the Deutsche Forschungsgemeinschaft (DFG,
German Research Foundation) - 314838170, GRK 2297 MathCoRe.

\printbibliography

\end{document}